\documentclass[%
 aip,
 amsmath,amssymb,
 reprint,%
]{revtex4-1}

\usepackage{graphicx}
\usepackage{dcolumn}
\usepackage{bm}

\usepackage[utf8]{inputenc}
\usepackage[T1]{fontenc}
\usepackage{mathptmx}
\usepackage{etoolbox}
\usepackage{soul,color}

\makeatletter
\def\@email#1#2{%
 \endgroup
 \patchcmd{\titleblock@produce}
  {\frontmatter@RRAPformat}
  {\frontmatter@RRAPformat{\produce@RRAP{*#1\href{mailto:#2}{#2}}}\frontmatter@RRAPformat}
  {}{}
}%
\makeatother
\begin{document}

\preprint{AIP/123-QED}

\title[On the solvable-unsolvable transition
due to noise-induced chaos in digital memcomputing]{On the solvable-unsolvable transition
due to noise-induced chaos in digital memcomputing}
\author{D. C. Nguyen}
\affiliation{
Department of Physics and Astronomy, University of South Carolina,
Columbia, SC 29208, USA
}%

\author{T. Chetaille}
\affiliation{
{\'E}cole Polytechnique F{\'e}d{\'e}rale de Lausanne, Lausanne, 1015, Switzerland
}%

\author{Y.-H. Zhang}
\affiliation{
Department of Physics, University of California, San Diego, La Jolla, CA 92093-0319, USA
}%

\author{Y. V. Pershin\textcolor{black}{$^*$}}
\email{\textcolor{black}{pershin@physics.sc.edu}}
\affiliation{
Department of Physics and Astronomy, University of South Carolina,
Columbia, SC 29208, USA
}%

\author{M. Di Ventra}
\affiliation{
Department of Physics, University of California, San Diego, La Jolla, CA 92093-0319, USA
}%

\date{\today}

\begin{abstract}
Digital memcomputing machines (DMMs) have been designed to solve complex combinatorial optimization problems. Since DMMs are fundamentally classical dynamical systems, their ordinary differential equations (ODEs) can be efficiently simulated on modern computers. This provides a unique platform to study their performance under various conditions. An aspect that has received little attention so far is how their performance is affected by the numerical errors in the solution of their ODEs and the physical noise they would be naturally subject to if built in hardware. Here, we analyze these two aspects in detail by varying the integration time step (numerical noise) and adding stochastic perturbations (physical noise) into the equations of DMMs. We are particularly interested in understanding  how noise induces a chaotic transition that marks the shift from successful problem-solving to failure in these systems. Our study includes an analysis of power spectra and Lyapunov exponents depending on the noise strength. The results reveal a correlation between the instance solvability and the sign of the ensemble averaged mean largest Lyapunov exponent. Interestingly, we find a regime in which DMMs with positive mean largest Lyapunov exponents still exhibit solvability. Furthermore, the power spectra provide additional information about our system by distinguishing between regular behavior (peaks) and chaotic behavior (broadband spectrum). Therefore, power spectra could be utilized to control whether a DMM operates in the optimal dynamical regime. Overall, we find that the qualitative effects of numerical and physical noise are mostly similar, despite their fundamentally different origin.
\end{abstract}

\maketitle

\begin{quotation}

Developing efficient methods to solve challenging optimization problems is crucial in various technological fields, including logistics, manufacturing, and cryptography. In recent years, significant progress has been made in the area of digital memcomputing -- a promising approach to tackle challenging optimization problems. Digital memcomputing machines (DMMs) solve problems by mapping them into dynamical systems and evolving these systems over time. However, in realistic \textcolor{black}{settings}, various factors, such as inaccuracies during numerical simulations or imperfections in hardware, introduce sources of noise that may impede the ability of DMMs to find solutions. In this paper, we show that with increasing noise strength, such noise sources can lead the system into a chaotic regime, where the DMM is unable to find the solution. Understanding this noise-induced solvable-unsolvable transition may help to create more stable and noise-tolerant dynamical systems designed to tackle complex problems.

\end{quotation}

\section{Introduction} \label{sec:1}

In the field of computing, a new class of non-linear (classical) dynamical systems has been recently introduced, designed to solve complex combinatorial optimization problems~\cite{Traversa17a}. These dynamical systems have been named digital memcomputing machines (DMMs) in view of the fact that they map a finite string of symbols to a finite string of symbols while coupling two types of degrees of freedom: ``fast ones'' representing the variables of the problem to solve, and ``slow ones'' representing memory degrees of freedom~\cite{MemComputingbook}. Since DMMs are classical systems, their ordinary differential equations (ODEs) can be efficiently simulated on our modern computers. In fact, most of the studies concerning DMMs so far have been performed by simulating their ODEs~\cite{traversa2017polynomial,traversa2018evidence,traversa2018memcomputing,sheldon2019stress,Sean3SAT,YuanhangDP,sharp2023scaling}\footnote{Also see, e.g., case studies in~\cite{MemComputingbook} or those reported by the company MemComputing, Inc.: www.memcpu.com.}, with a few implementations on small and large FPGA boards~\cite{chung23,10658882}.

Despite much work in this fledgling field has focused on understanding the topological aspects of these dynamical systems~\cite{topo,di2019DLRO}, less attention has been devoted to the role of noise, whether in the form of numerical errors introduced in the discretization of the corresponding ODEs, or the physical noise that would naturally appear in a
hardware realization of these machines. {\it Numerical noise} can be increased during the time discretization process, by increasing the integration time step, resulting in discrete approximations that alter the path of the system in phase space. {\it Physical noise} occurs due to variations in the environment or control parameters, adding complexity and unpredictability.
Although there have been multiple studies investigating noise effects in discrete maps and Hamiltonian systems \cite{moss1989noise,nychka1992finding}, the influence of noise on the transition to chaos within continuous-time dynamical systems remains less examined~\cite{chaosororder}. \textcolor{black}{Nevertheless, the phenomenon of noise-induced chaos has been recognized for several decades. Early works \cite{iansiti1985noise, Bulsara1990} revealed how noise can drive trajectories into fractal basin boundaries, producing switching-like behavior. More recently \cite{Lai2003, Tel2008}, it has been shown that transient chaos, which is present in noise-free systems can, under sufficiently strong noise, become sustained chaos, thus giving rise to chaotic attractors.}

The goal of this paper is to precisely investigate the effect of both types of noise, with the physical noise introduced as stochastic perturbations into the equations of DMMs, while keeping the numerical noise under control. In particular, we are interested in understanding the noise-induced transition from the ability of these machines to solve problem instances to the failure to do so. We will show that this noise-induced transition is a manifestation of noise-induced chaos.

Here, we examine the noise-induced transition from a solvable to an unsolvable phase in prototypical DMMs as those introduced in~\cite{Sean3SAT}. Such DMMs have been designed to solve 3-SAT instances (a collection of logical disjunctions of exactly three variables or their negations, related to each other by logical conjunctions), a well-known combinatorial problem~\cite{complexity-bible}.
These DMMs solve these instances efficiently thanks to the emergence of \textit{dynamical long-range order} (DLRO)~\cite{di2019DLRO} and the fact that, if properly designed, they admit only saddle points and equilibria corresponding to the solution(s) of the problem instance. No other local minima are present in the phase space.

In the language of field theory, the DLRO of DMMs is expressed by the collective behavior of specific trajectories in the phase space, called instantons, or avalanches~\cite{di2019DLRO}, which connect critical points of the flow vector field of the dynamics with decreasing index, namely from saddle points with a certain number of unstable directions to saddle points with less unstable directions. We refer to the book~\cite{MemComputingbook} for a thorough explanation of these statements and to Ref.~\cite{Sean3SAT} for their
mathematical demonstration. The very fact that DMMs employ critical points (saddle points) and trajectories between them (instantons) make them topologically robust, in the sense that their dynamics are fairly insensitive to perturbations. However, we expect this robustness to reach a limit as a function of the noise strength introduced during dynamics, whether numerical or physical.

In this study, by varying the noise strength, we demonstrate the following common tendency for both numerical and physical noise: A progression from {\it i)} non-chaotic dynamics to {\it ii)} ``transiently chaotic'' (in the sense of a small but positive mean largest Lyapunov exponent), followed by {\it iii)} chaotic behavior and potentially leading to {\it iv)} a (quasi-)periodic-like dynamic regime (only in the case of numerical noise).  Through extensive simulations, we analyze Lyapunov exponents and power spectra, to gain insights into some of their characteristics. While {\it iii)} and {\it iv)} obviously lead to a breakdown of the DMMs' ability to solve problem instances, we find that regime {\it ii)} still provides solvability, further reinforcing the notion that DMMs are topologically robust.

\textcolor{black}{In the context of nonlinear dynamics, the term ``transient chaos'' traditionally refers to chaotic behavior that persists only for a finite time before the system evolves towards a fixed point or a limit cycle~\cite{Lai2011}. This behavior arises due to the presence of a non-attracting chaotic invariant set, chaotic saddles, in phase space, near which the trajectories evolve chaotically before escaping.
However, our focus is on the trajectory that leads to the solution, and we stop the dynamics once the solution is reached and do not analyze the subsequent asymptotic regime. In our system, we can visualize the existence of transient chaos by examining the number of steps until solution as a function of the initial condition (we show this explicitly in the Appendix~\ref{Appendix_tran}). The sensitive non-monotonic dependence we observe is analogous to the results that show the characteristic signature of transient chaos in Ref.~\cite{Lai2011}.}

The structure of this paper is as follows. Sec.~\ref{sec:2} describes the methods employed in this investigation. In Sec.~\ref{sec:3}, we detail the modeling results, beginning with the case of numerical noise (Sec.~\ref{sec:3a}) and subsequently addressing the case of physical noise (Sec.~\ref{sec:3b}). We discuss our results in Sec.~\ref{sec:4}, and conclude in Sec.~\ref{sec:5}.

\section{Methods} \label{sec:2}

\subsection{DMMs equations for 3-SAT}

Let us consider a Boolean satisfiability problem referred to as 3-SAT, which comprises $N$ Boolean variables and $M$ clauses. Each clause is a disjunction of three literals (variables or their negation). \textcolor{black}{Formally, a clause $m$ can be defined by a set $\{\{p^m_1,q^m_1\},\{p^m_2,q^m_2\},\{p^m_3,q^m_3\}\}$, where $p^m_i$ is the index of the variable entering the clause $m$ at position $i=1,2,3$, and $q^m_i=\mp1$ defines whether the variable $\textnormal{v}_{p^m_i}$ is negated or not, respectively.} A solution to the 3-SAT instance involves assigning values to these variables so that every clause evaluates to 1 (TRUE). Since our goal for this paper is to understand the role of noise in the ability of DMMs to solve the problem, we have generated instances following Ref.~\cite{barthel2002hiding} so that we can guarantee that they have solutions.

A DMM designed to solve such a problem would use some type of dissipative dynamical systems, where solutions to the 3-SAT correspond to stable equilibrium points of the dynamics (see, e.g., Ref.~\cite{Traversa17a}). However, the choice of the particular dynamical system representing such a problem is not unique. Here, we will consider the one reported in~\cite{Sean3SAT,pershin2024accurate} which has been shown to be able to tackle hard instances quite efficiently. These particular DMMs have ODEs of the form (the arguments $\textrm{v}_n,\textrm{v}_j,$ and $\textrm{v}_k$ are
variables appearing in a single clause of the 3-SAT instance):
\begin{widetext} \textcolor{black}{
\begin{eqnarray}
\dot{\textnormal{v}}_n&=&\sum\limits_{m=1}^My_{m}x_{m}G_{n,m}+\left( 1+\zeta y_{m}\right) \left(1-x_{m}\right) R_{n,m}, \hspace{3.1cm} n=1,\ldots,N, \label{eq:1}\\
\dot{x}_{m}&=&\beta \left( x_{m} +\epsilon \right)\left( C_m-\gamma\right), \hspace{6.45cm} m=1,\ldots,M, \label{eq:2}\\
\dot{y}_{m}&=&\alpha \left( C_m-\delta\right), \hspace{7.8cm}  m=1,\ldots,M, \label{eq:3}\\
C_m&=&\frac{1}{2}\text{min}\left[\left(1-q^m_1\textnormal{v}_{p^m_1} \right),\left(1-q^m_2\textnormal{v}_{p^m_2} \right),\left(1-q^m_3\textnormal{v}_{p_3^m} \right)\right],\label{eq:6}
\\
G_{n,m}&=&\begin{cases}
 0, \hspace{7mm} n\notin \{p^m_1,p^m_2,p^m_3\},\\
 \frac{1}{2} q^m_i\text{min}\left[\left( 1-q^m_j\textnormal{v}_{p^m_j}\right), \left( 1-q^m_k\textnormal{v}_{p^m_k}\right)\right]\mid \{ p^m_i=n, j\neq k\neq i\}, \hspace{1cm }\text{otherwise},
\end{cases}
\\
R_{n,m}&=&\begin{cases}
    q^m_iC_m
    \hspace{1cm} \text{if } C_m=\frac{1}{2}\left(1-q^m_i\textnormal{v}_{p^m_i} \right)\mid \{ p^m_i=n\} \\
    0, \hspace{7mm} \text{otherwise}. \label{eq:5}
  \end{cases}
\end{eqnarray}}
\end{widetext}
Here, \textcolor{black}{ $N$ is the number of variables, $M$ the number of clauses and }$\textnormal{v}_n\in [-1,1]$ (with $1\leq n \leq N$) are continuous versions of Boolean variables. In an electrical circuit realization of these DMMs, the variables would be, e.g., the voltages at the Boolean gate terminals~\cite{Traversa17a}. Irrespective, we consider $\textnormal{v}_n >0$ to correspond to TRUE and $\textnormal{v}_n < 0$ to FALSE. The variables, $y_{m}$ and $x_{m}$ are long- and short-memory variables bounded in $[1,10^4 M]$ and $[0,1]$, respectively, whose dynamics are slower than those of the $\textnormal{v}_n$'s. \textcolor{black}{The upper bound of $y_m$ was arbitrarily imposed to guarantee compactness of the phase space. This bound is high enough to prevent the long-term memory from reaching that ceiling. }
Additionally, $\alpha$, $\beta$, $\gamma$, $\delta$, $\epsilon$ and $\zeta$ are constants. Following~\cite{Sean3SAT}, we use the following set of parameter values (tuned for the instances of Ref.~\cite{barthel2002hiding}): $\alpha = 5$, $\beta = 20$, $\delta = 0.25$, $\gamma = 0.05$, $\epsilon = 10^{-3}$, $\zeta = 0.1$ (for $M/N=7$), and $\zeta = 0.001$ (for $M/N=4.3$).

Equations~(\ref{eq:1})-(\ref{eq:3}) were integrated numerically using the forward Euler method \cite{NumRecipes} and added noise to emulate physical noise (see also Secs.~\ref{physnoise},~\ref{sec:3a} and~\ref{sec:3b} below for more details).
The clause function \textcolor{black}{ $C_m(\textnormal{v}_{p_1^m},\textnormal{v}_{p_2^m},\textnormal{v}_{p_3^m})$} is designed such that $C_m < 0.5$ if the clause is satisfied. Therefore, the entire instance is solved when $C_m < 0.5$ for all $m=1,\ldots,M$. It has been shown~\cite{Sean3SAT} that for all solvable instances, Eqs.~(\ref{eq:1})-(\ref{eq:3}) converge to a solution.

\subsection{Lyapunov exponent and power spectrum analyses}
{\it Lyapunov exponents --} In our study we determine the Lyapunov exponents to characterize the dynamics of DMMs. In the literature, a positive largest Lyapunov exponent (LLE) is typically used as an indicator of chaotic behavior~\cite{Hilborn}.
The usual technique for calculating the LLE is the simultaneous evolution of one perturbation vector along with the system trajectory, with periodic renormalization to keep the norm fixed~\cite{ispolatov2015chaos,hochstetter2021avalanches,aurell1997predictability,zoltan}. However, since the variables in Eqs.~(\ref{eq:1})-(\ref{eq:3}) are bounded, this can be problematic because the orientation of maximum expansion may be outside of the phase space. If this occurs, just rescaling the perturbation along this direction would lead to unphysical states.


To avoid this problem, in the study of numerical noise, we implemented an algorithm introduced by Benettin {\it et al.}~\cite{benettin1980lyapunov}. The Benettin method avoids this problem by generating a basis of perturbation vectors spanning the tangent space, applying Gram-Schmidt orthonormalization at each step to ensure they remain independent and normalized. This approach eliminates the necessity for reorientation of perturbations, allows the system to select acceptable directions, and maintains all computed exponents physically meaningful in the phase space.

We compute the entire Lyapunov spectrum by slightly perturbing the trajectory in $N$ orthogonal directions with a small perturbation $\Delta \mathbf{\textrm{v}}_{0,k}$ (in the $k$-direction) and integrating each with identical time steps $\Delta t$. We compute the \textcolor{black}{ instantaneous } Lyapunov  exponents as
\begin{equation}
    \lambda_k = \frac{1}{\Delta t} \text{ln} \left(\frac{||\Delta \mathbf{v}_k||}{||\Delta \mathbf{v}_{0,k}||}\right),
    \label{lyapunov_formula}
\end{equation}
\textcolor{black}{ where $\|\cdot\|$ denotes the Euclidean norm of the perturbation vector }.
At each time step, the perturbations were orthogonalized using the modified Gram-Schmidt orthonormalization process~\cite{bjorck1992loss,bjorck1994numerics} and normalized. One then chooses the largest \textcolor{black}{ instantaneous } Lyapunov exponent denoted by $\lambda$ at each time step. In what follows, the following notation is used: $\overline{\lambda}$ for the mean largest Lyapunov exponent (MLLE; time-averaged over one specific trajectory from $t=0$ to the smallest of \{solution time, maximum evolution time\}), and $\langle\overline{\lambda}\rangle$ for the ensemble averaged mean largest Lyapunov exponent (EAMLLE), where additional averaging occurs over instances of the problem.

\textcolor{black}{We note that in the noise-free case the system always converges to the point attractors corresponding to the system solutions, and therefore the asymptotic LLE is zero. In this work, we simulated the equations and computed LLE up to the time when the solution was found and thus should be interpreted as finite time LLE. For sufficiently long times, especially in the weak noise regime, LLE would of course decay to zero. This distinction is consistent with the standard definition of noise-induced chaos~\cite{Lai2003,Lai2011}, where the long term exponent remains zero below a critical noise strength and becomes positive only above it. }

In the case of physical noise, the evaluation of the largest Lyapunov exponent involved removing the restriction on the values of $\textnormal{v}_n$ (refer to the line under Eq.~(\ref{eq:6})), as discussed in~\cite{pershin2024accurate}. This removal enabled us to apply the conventional method for calculating the LLE. However, the gradient norm of $\textnormal{v}_n$
is capped at $10^4$ to avoid instability and prevent failure due to unbounded variables and excessively large gradient updates. Additional information can be found in Sec.~\ref{sec:3b} below.

{\it Power spectra --} To complement the Lyapunov exponent and have a better understanding of the DMM's dynamics, we also calculated the power spectrum of the fast variables' flow vector field $\textnormal{d}\mathbf{v}/\textnormal{d}t=(\dot{\textnormal{v}}_1, \dot{\textnormal{v}}_2, .... \dot{\textnormal{v}}_N)$ to discern the presence of dominant frequencies or a broadband spectrum characteristic of chaos.

For a signal $x(t)$, its power spectrum is usually given by the modulus square of the Fourier components per unit time~\cite{vanKampen}:
\begin{equation}
    \label{psepctrum}
    S(f)=\lim_{T\to\infty} \frac{1}{T} \bigg\rvert\int_0^Tx(t)e^{-j2\pi ft}\bigg\lvert ^2
\end{equation}
For discrete time signals $x[n]$ of length $N$, the Fast Fourier Transform yields $S[k]=\frac{1}{N f_s}\rvert X[k]\lvert^2$, where
$X[k]=\sum_{n=0}^{N-1}x[n]e^{-j\pi n/N}$ and $f_s=\frac{1}{\Delta t}$~\cite{pspectrumcomputation}. The power spectrum is obtained by performing this calculation on the right-hand side of Eq.~(\ref{eq:1}) for each variable independently, and then averaged by the number of variables $N$. \textcolor{black}{Therefore $x(t)$ in Eq. (\ref{psepctrum}) would represent here a component of $\textnormal{d}\mathbf{v}/\textnormal{d}t$.}

Note that we have opted to consider the power spectrum of the flow vector field, rather than of the $\textnormal{v}_n(t)$'s to avoid a zero-frequency peak associated with the stable variables' values $\pm 1$. These fast degrees of freedom are the primary dynamic variables whose evolution determines the trajectory in the solution space of the 3-SAT problem. The memory variables, while essential for shaping the dynamics (e.g., inducing transitions or stabilizing solutions), act as ``modulators''. Consequently, their time series contain little information about the unstable transitions characteristic of chaos.

\subsection{Physical noise} \label{physnoise}
To make contact with experimental realizations of these machines, we introduce physical noise through stochastic terms in the ODEs of DMMs. In view of the fact that in possible electrical circuit realizations of DMMs, the memory variables are the ones most susceptible to noise~\cite{Traversa17a}, we add stochastic terms to Eqs. (\ref{eq:2}) and (\ref{eq:3}). Of course, such noise propagates indirectly to the fast degrees of freedom $\textnormal{v}_n$'s through their coupling in Eq. (\ref{eq:1}). Indeed, as discussed in \cite{Bearden_2018}, noise can, e.g., serve as a proxy for temperature, since temperature fluctuations can induce physical instabilities in the resistive memory elements of the gates of DMMs.

If we indicate with $\tilde{y}_{m}$ the noisy version of the long-term memory variables, the stochastic terms are linked to the noiseless dynamical system's memory by
\begin{equation}
    \label{stochastic terms}
    \dot{\tilde{y}}_{m}(t)=\dot y_{m}(t)+\Gamma \eta(t)
\end{equation}
with $\eta \sim \mathcal{N}(0,1)$, a Gaussian distribution with zero average and unit variance, and $\Gamma$ is the strength of the noise. Similarly for the short term memory $x_{m}$. 

\begin{figure}[t]
    \includegraphics[width=0.85\columnwidth]{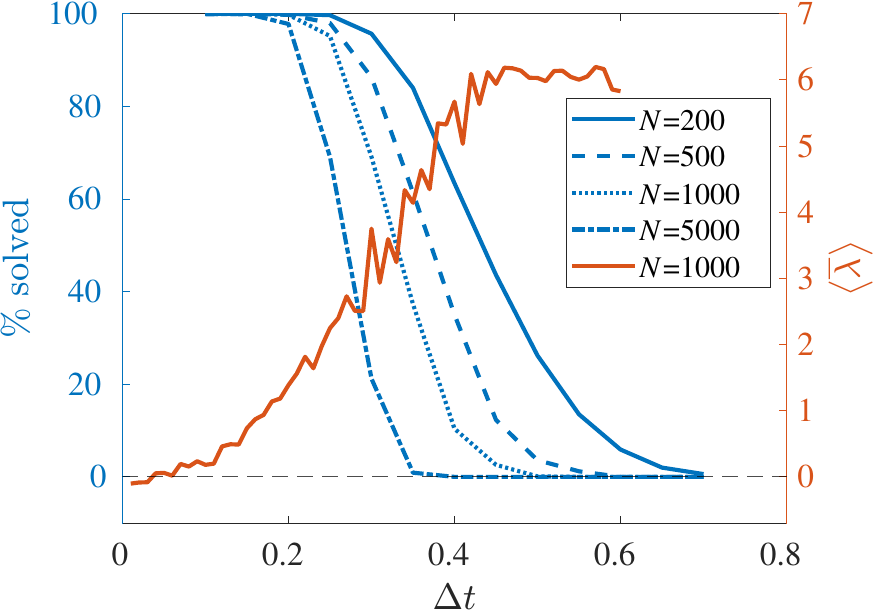}%
    \caption{\textcolor{black} {Numerical noise investigation:}  Percentage of solved instances and EAMLLE ($\langle\overline{\lambda}\rangle$) vs. integration time step, $\Delta t$. Each point used to make this plot was obtained using an ensemble of 100 solvable 3-SAT instances with clause-to-variable ratio $M/N=7$. The upper bound for each trajectory was set to 20,000 steps. \textcolor{black}{ The smallest $\Delta t$ taken is 0.01.} }
    \label{fig:1}
\end{figure}


\section{Results} \label{sec:3}

\subsection{Numerical noise} \label{sec:3a}

By numerically integrating Eqs.~(\ref{eq:1})-(\ref{eq:3}) using the forward Euler method, we transform the continuous-time dynamics into a discrete map. We view related errors induced by such a transformation as the primary components of numerical noise, which can sometimes force the system into chaotic behavior.
It is clear that the solution becomes unattainable when the time step is too large. In fact, numerical noise typically accumulates during the integration of the ODEs~\cite{NumRecipes}, hence it is ``non-local'' in time. This is quite distinct from the physical noise in Eq.~(\ref{stochastic terms}) which is instead local in time.

\begin{figure*}[tb]
\centering
    (a) \includegraphics[width=0.75\columnwidth]{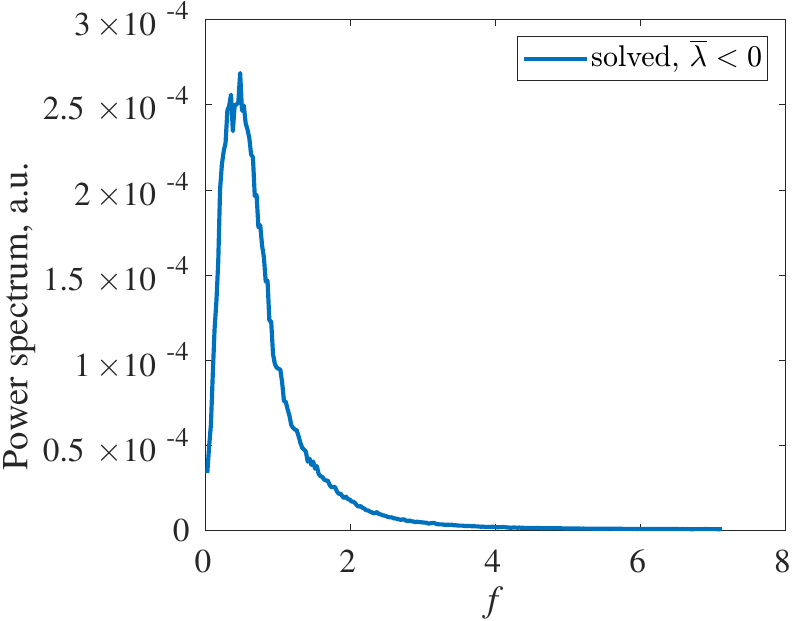}
    (b) \includegraphics[width=0.75\columnwidth] {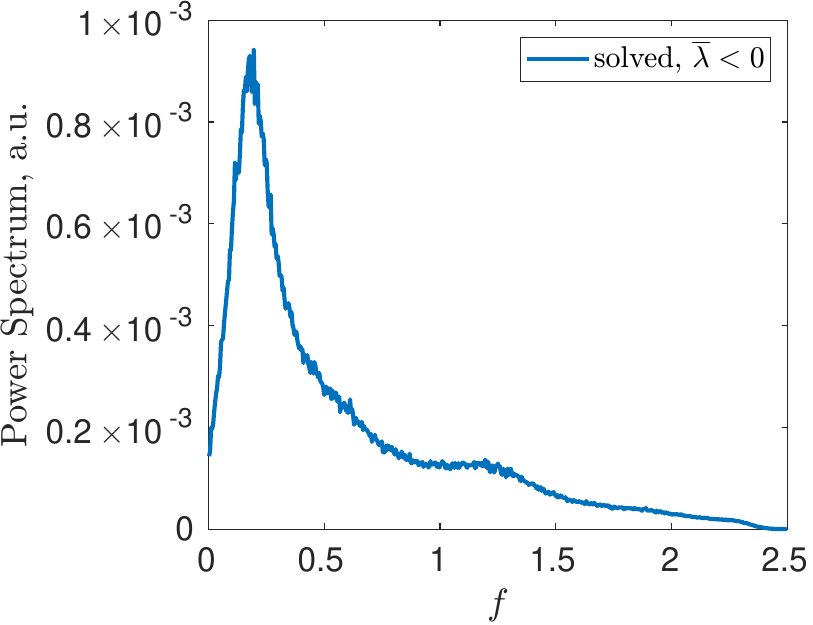} \\
    (c) \includegraphics[width=0.75\columnwidth]{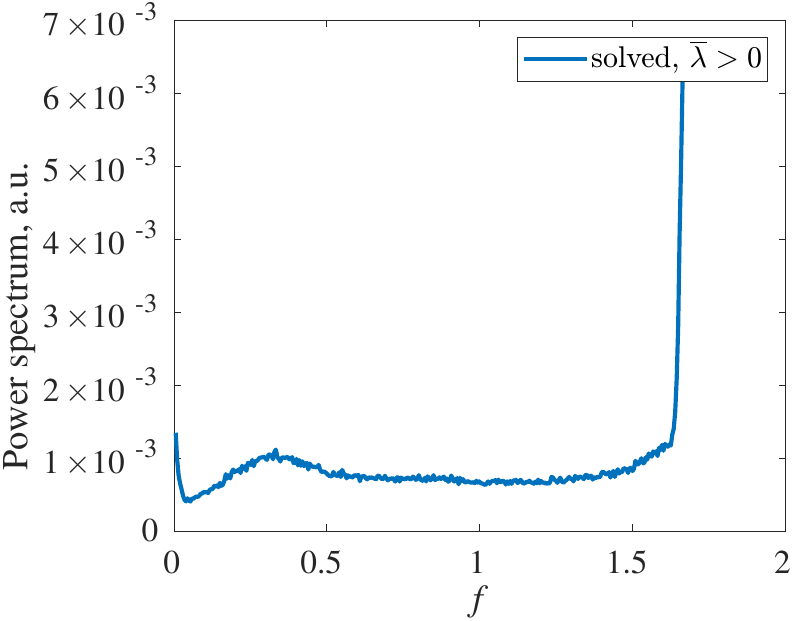}
    (d) \includegraphics[width=0.75\columnwidth]{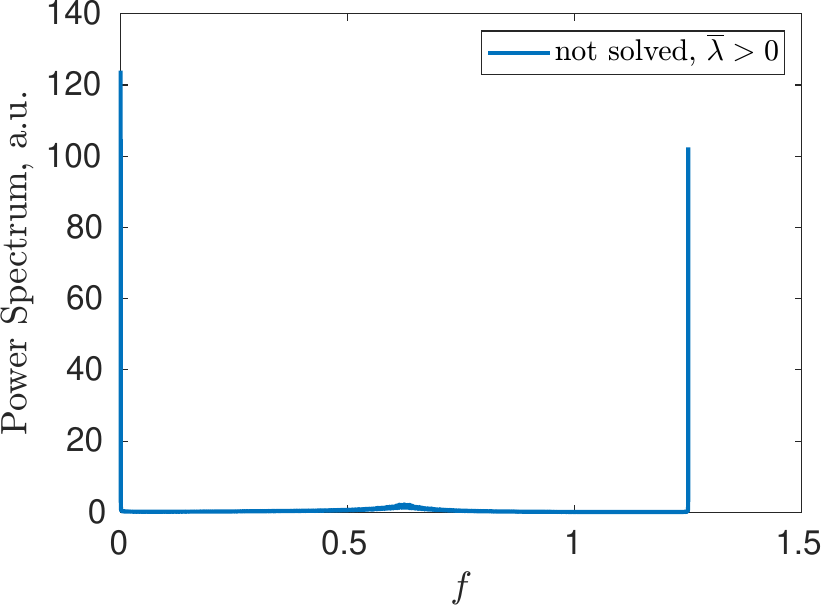}
    \caption{\textcolor{black}{ Numerical noise investigation:} Power spectra for a satisfiable instance with $N=1000$ variables and  with clause-to-variable ratio $M/N=7$ generated using the integration time step of (a) 0.07, (b) 0.2, (c) 0.3, and (d) 0.4.}
    \label{fig:2}
\end{figure*}

Fig.~\ref{fig:1} illustrates how the percentage of solved instances and the EAMLLE vary with integration time step in problems with clause-to-variable ratio $M/N = 7$.  In these simulations, the system's dynamics continued until a solution was reached or the simulation hit the maximum allowed number of steps.
\textcolor{black}{We note that the total time of integration varies according to the instances to allow enough time to reach a solution of the 3-SAT problem. }
As expected, one observes that with an increase in the integration time step, the proportion of solved instances drops from 100\% to 0\%.  The position of this drop correlates with the difficulty of the problem (characterized by the number of variables $N$): the solvable-unsolvable transition becomes sharper as the number of variables, $N$, increases. At this stage, however, we cannot conclude that such a transition is continuous in the
``thermodynamic limit'' ($N\rightarrow\infty$) or will become discontinuous (first order), since this distinction would require a theoretical analysis which is beyond the scope of this paper. However, we note that similar trends were observed in a recent paper~\cite{YuanhangDP}, and in Sec.~\ref{adaptive} we will show a similar phenomenon in the presence of physical noise.

According to our simulations, for the smallest $\Delta t$, the EAMLLE is negative across most instances. Fig.~\ref{fig:1} shows that the EAMLLE increases monotonically with $\Delta t$. However, there is an interesting region, $0.05 \lesssim \Delta t \lesssim 0.2$, where the positive EAMLLE co-exists with the 100~\% solvability. We associate this region with a ``transiently chaotic'' dynamics regime. Appendix \ref{AppendixA} shows LLE examples for two different time-step values.


\begin{figure*}[tb]
\centering
    (a) \includegraphics[width=0.75\columnwidth]{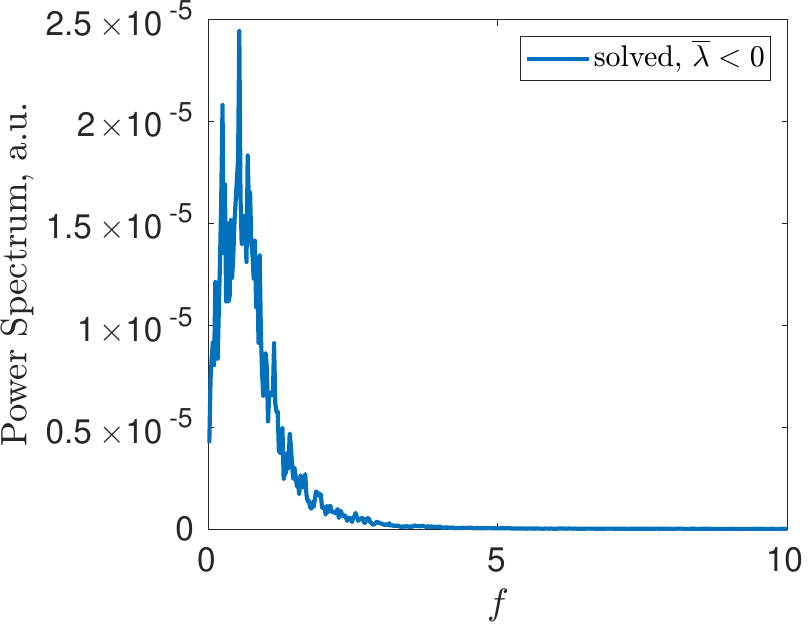}
    (b) \includegraphics[width=0.75\columnwidth]{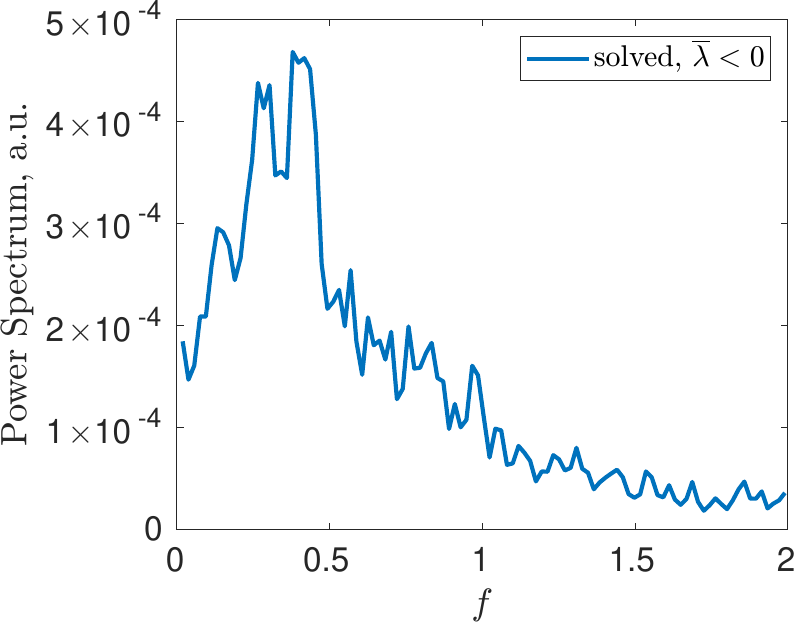}\\
    (c) \includegraphics[width=0.75\columnwidth]{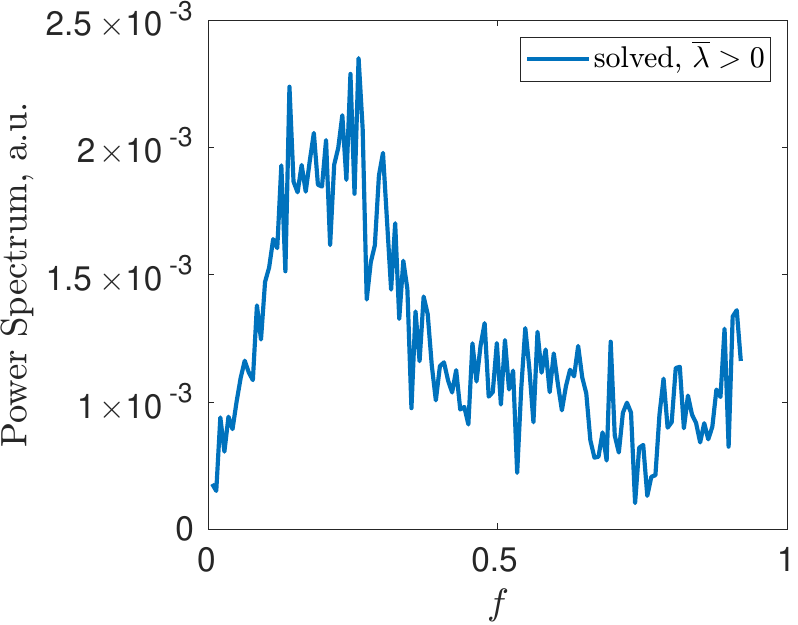}
    (d) \includegraphics[width=0.75\columnwidth]{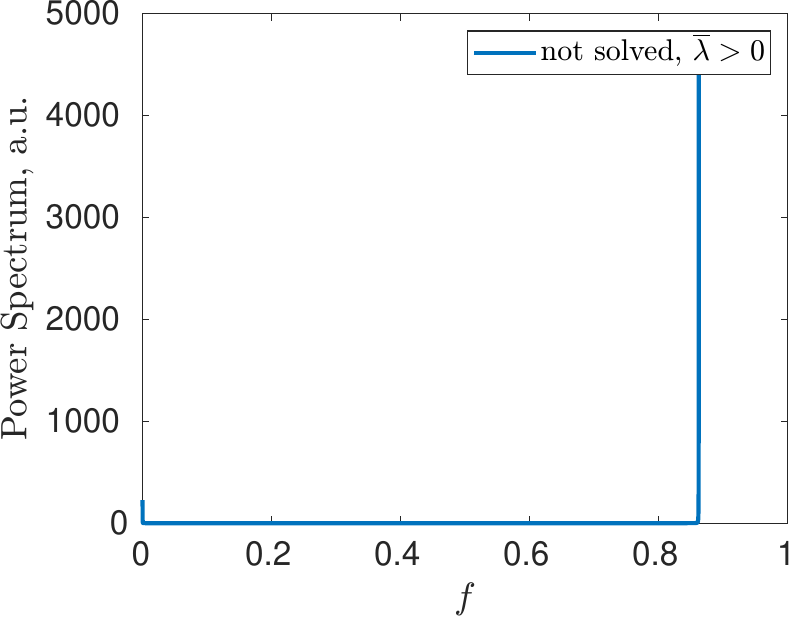}
    \caption{\textcolor{black}{ Numerical noise investigation: } Power spectra for a satisfiable instance with $N=90$ variables and  with clause-to-variable ratio $M/N=4.3$ obtained using the integration time step of (a) 0.05, (b) 0.25, (c) 0.54, and (d) 0.58.}
    \label{fig:3}
\end{figure*}

Fig.~\ref{fig:2} shows an example of the power spectra for a typical instance. For $\Delta t=0.07$ we observe a prominent peak at lower frequencies (Fig.~\ref{fig:2}(a)), which gradually diminishes at larger time steps. We defer to Sec.~\ref{peaks} for the explanation of the origin of this peak since it does not involve noise (whether numerical or physical). Here, we just focus on the numerical noise effects. As the time step increases, the system ceases to reach a solution and becomes trapped in a possibly quasi-periodic orbit according to Fig.~\ref{fig:2}(d). In the Supplementary Information (SI), we provide power spectra for two other instances of the same size, see Figs.~S1 and S2. 

Similar results are obtained for other types of instances. For example, in Fig.~\ref{fig:3} we present typical power spectra for
instances with clause-to-variable ratio $M/N=4.3$, which are considered to be much harder to solve compared to $M/N=7$ instances~\cite{Sean3SAT}. A similar peak at low frequencies is observed for small $\Delta t$, which decreases as $\Delta t$ increases. The primary distinction for large instances is that, when $\Delta t$ is sufficiently large, the instance remains unsolved yet does not stall at the boundaries, there is a significant peak at zero frequency and a nearly flat high-frequency band that gradually increases, as shown on the inset of Fig.~S1(d). Compared to Fig.~\ref{fig:2}, the curves in Fig.~\ref{fig:3} are also noisier confirming that the instances are harder.

\subsection{Physical noise} \label{sec:3b}

\begin{figure*}[t]
\centering
    \includegraphics[width=12cm]{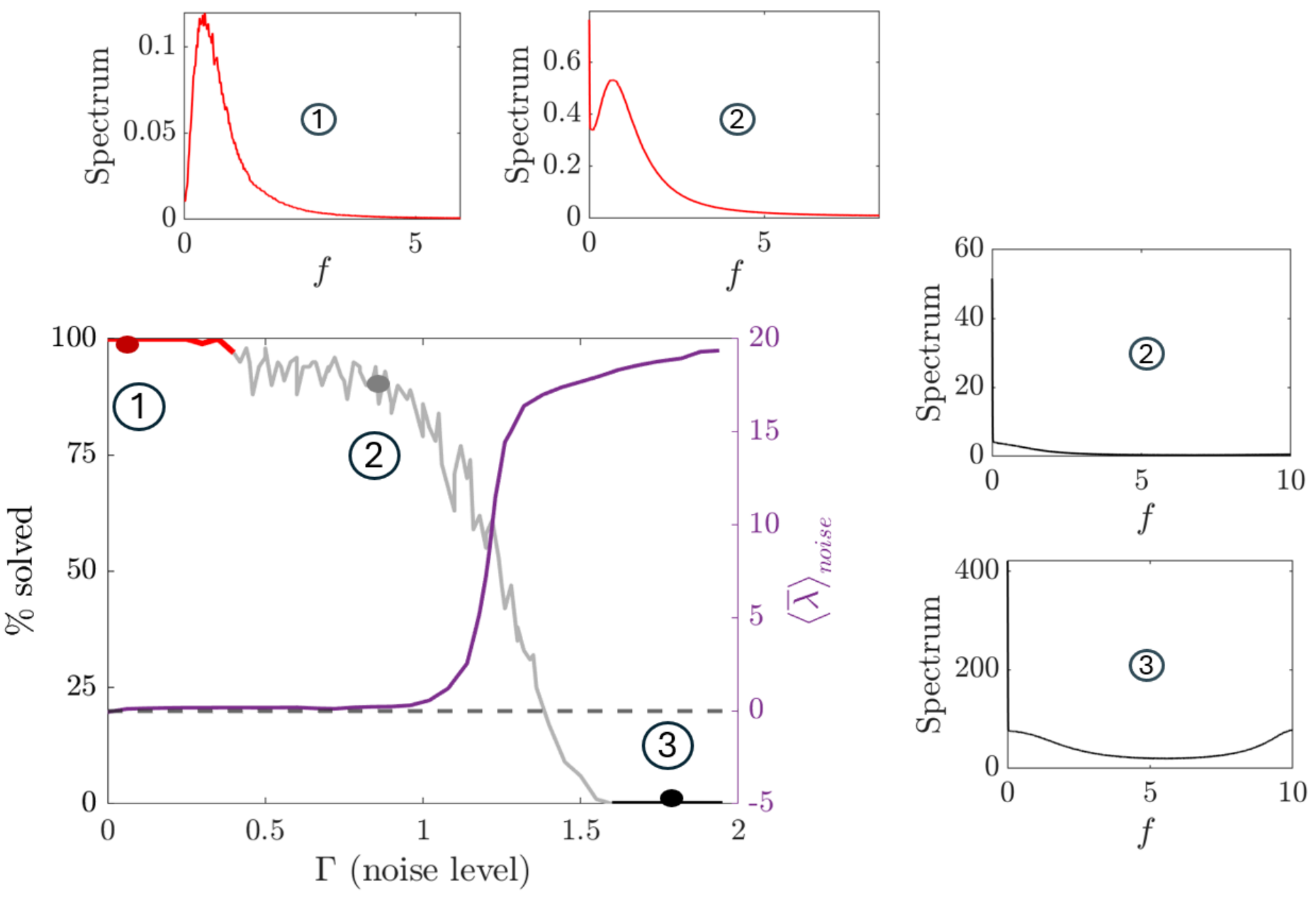}
    \caption{\textcolor{black}{ Physical noise investigation. } Main panel: Transition of the percentage of solved cases as a function of noise amplitude $\Gamma$, for a problem instance of $N=1000$ variables at clause-to-variable ratio $M/N=7$. Each point represents 100 realization of the noise. \textcolor{black}{The simulation runs for a maximum of 10,000 steps if convergence to a solution does not occur beforehand }. In all cases, the variables $\textnormal{v}_n$'s are initialized to zero, the long term memory $y_{m}$ to 1 and the short term memory $x_{m}$ to $C_m$ which is equal to 1/2 when $\textnormal{v}_n=0$ $\forall n$ (midpoint value of the short term memory bounded range [0,1]). The spectrum is then averaged over 100 noise realizations. The MLLE averaged over noise realisation $\langle\overline{\lambda}\rangle_{noise}$ is also displayed. Insets (1), (2), and (3) show representative power spectra of the flow vector field of the fast degrees of freedom of individual instances at different noise levels.
Inset (1): Power spectrum of a solved instance at low noise ($\Gamma \approx 0$), where 100\% of instances are solved.
Inset (2): Power spectra at intermediate noise. The red curve corresponds to the average power spectrum of noise realizations of 85 solved cases, and the black curve to the average of noise realizations of 15 unsolved ones. Inset (3): Average power spectrum of an unsolved instance at high noise over 100 noise realizations ($\Gamma \approx 1.7$ where the solving rate drops to zero).}
    \label{transition_n=1000}
\end{figure*}

We now move on to discuss the effect of physical noise as introduced according to the procedure discussed in Sec.~\ref{physnoise}. We then initialize all fast variables $\textnormal{v}_n$'s to zero and increase the noise strength $\Gamma$. This is illustrated in Fig. \ref{transition_n=1000}, where the percentage of solved cases for a single problem instance (with $N=1000$ variables)  with respect to the noise strength is plotted for clause-to-variable ratio $M/N=7$, along with the power spectrum of the fast variables' flow vector field $\textnormal{d}\mathbf{v}/\textnormal{d}t$ for representative instances. A problem is considered unsolved if it does not reach a solution within $10^4$ steps.
\textcolor{black}{ The physical noise introduced in Eq. \eqref{stochastic terms} is genuinely stochastic, so each trajectory corresponds to a distinct realization of the random process $\eta(t)$, while the problem instance and initial conditions are kept fixed. Ensemble-averaged quantities such as $\langle \overline{\lambda }\rangle_{\text{noise}}$ are therefore obtained by averaging over 100 independent noise realizations. In contrast, the numerical noise in Sec.~III~A is deterministic once $\Delta t$ is chosen, so there is no concept of ``realizations,'' and averaging is instead performed over different problem instances.}
The choice of initializing all the instances identically is justified in Fig. S6, which demonstrates that using random initial conditions and instances results in the same power spectrum behavior as using fixed ones. To minimize numerical errors, following the results presented in Fig.~\ref{fig:1}, we have chosen an integration time step $\Delta t$ of 0.05. (However, in Sec.~\ref{adaptive}, we consider an adaptive time step to further control the numerical errors while increasing the problem size.) The main plot of Fig. \ref{transition_n=1000} shows the sharp decline in the success rate as noise increases, indicating the solvable-to-unsolvable transition.

As for the case of numerical noise, we also expect in this case that the power spectrum of the vector signal $\textnormal{d}\textbf{v}/\textnormal{d}t$ should have large discrete peaks at particular frequencies \cite{Sydney}. However, the spacing between neighboring frequency components should become negligible for aperiodic sequences like chaotic signals \cite{pspsectrumchaos1, pspectrumchaos2}. The power spectrum of a chaotic solution should then contain all frequency components and to be broadband
continuous without distinct periodicity \cite{LIPTON1996290}.
This results in a spectrum that spreads continuously across frequencies and does not have any spikes.


In agreement with the results obtained with numerical noise (Fig.~\ref{fig:1}), Fig~\ref{transition_n=1000} shows that the physical noise induces a decay in the percentage of solved cases. We again interpret this decay as the appearance of noise-induced chaos.
Inset (1) of Fig~\ref{transition_n=1000} shows the power spectrum of the flow vector field $\textnormal{d}\mathbf{v}/\textnormal{d}t$ with no noise. There is a clear peak at low frequencies, showing a dominant frequency of the signal associated to regular dynamics.  Inset (2) in red shows the power spectrum for the solved cases with intermediate noise strength. The peak is still there, but a drop at low frequencies due to $1/f$ noise appears. (We refer to Appendix ~\ref{1/f_noise} for a discussion of this type of noise.) The spectrum is also smoother than in Inset (1) because an average has been computed over all noise realizations. The sub-harmonics have canceled themselves out, and only the main frequency component remains. For the unsolved cases in black, $1/f$ noise is still present for small frequencies, and for larger frequencies there is a continuum spectrum, indicating the presence of chaos. The same features are found in the Inset (3), when no cases are solved. The amplitude of the power spectrum increases again for large frequencies due to the long term memory.

Due to physical noise, the Lyapunov exponents are calculated using $\textnormal{v}_n$'s  that are not limited by fixed bounds as in Eq.~(\ref{eq:1}). This is necessary because with strong noise, both the original state and the small perturbation could get stuck at the edges of the system, making it impossible to measure their separation. However, to keep the system stable, the norm of $\textnormal{d}\mathbf{v}/\textnormal{d}t$ is limited to a maximum of $10^4$, so the $\textnormal{v}_n$'s do not grow indefinitely. The quantity $\langle \overline{\lambda}\rangle_{noise}$
is obtained by averaging the largest Lyapunov exponents at each time step, and then averaging again over different noise realizations. This quantity is plotted in the main panel of Fig.~\ref{transition_n=1000} and its trend is very similar to the one we already discussed for numerical noise (cf. Fig.~\ref{fig:1}).
\textcolor{black}{We emphasize again that the reported Lyapunov exponents correspond to finite-time, while the asymptotic value remains zero below a critical noise strength. According to the standard picture of noise-induced chaos, small stochastic perturbations act to sustain the transient chaos already present in the noise-free system. The noise continually kicks trajectories away from the regular attractor, allowing them to revisit the unstable region near the chaotic saddle. As the noise amplitude increases, trajectories spend progressively longer time near this saddle; beyond a critical noise strength, the random kicks prevent escape altogether, effectively transforming the chaotic saddle into a true chaotic attractor. This mechanism naturally accounts for the observed growth of finite-time Lyapunov exponents with increasing noise strength in our simulations. }
\subsection{Transition to chaos as a function of problem size} \label{adaptive}
To study the transition into chaos with respect to the system size, an adaptive time step had to be used to better control numerical errors. The inverse of the largest time derivative of $\textbf{v}$ is taken to scale the time step adaptively. This means that if the time derivative is large, the time step becomes small to prevent instability. Conversely, a smaller derivative leads to larger updates for faster progress. Therefore, $\Delta t$ is rescaled at every time step by a factor of $1/(\text{max}\rvert{\text{d}\textbf{v}}/{\text{d}t}\lvert +
\bar\varepsilon)$, with $\bar\varepsilon = 10^{-6}$  to avoid divergences when $\text{max}\rvert{\text{d}\textbf{v}}/{\text{d}t}\lvert=0$.

\begin{figure}[t]
\centering
    \includegraphics[width=0.85\columnwidth]{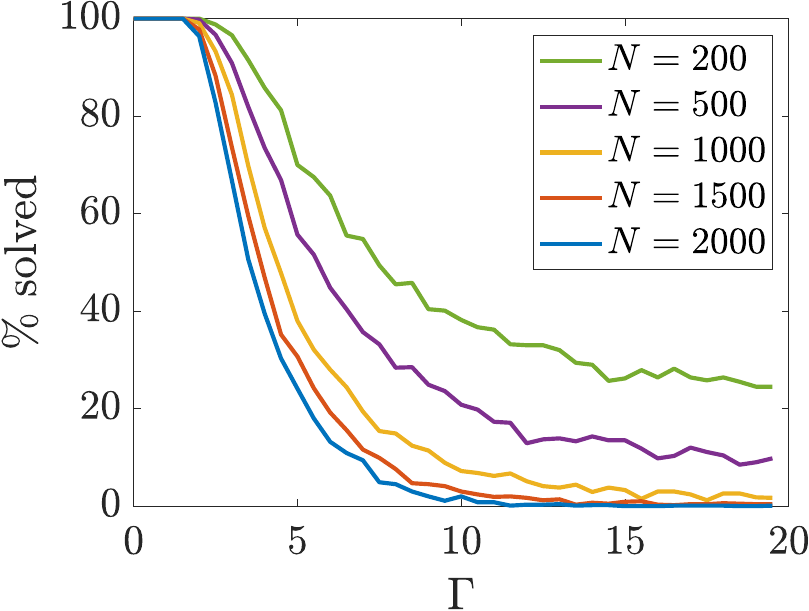}
    \caption{\textcolor{black}{ Physical noise investigation:} Percentage of problem instances solved with respect to the noise strength $\Gamma$ for clause-to-variable ratio $M/N=7$. The solvable-unsolvable transition becomes sharper with increasing number of variables. The time step is set to be inversely proportional to the maximum time derivative of the variables' vector $\textbf{v}$ (see main text for details).
    For every value of $\Gamma$ and for each size $N$ a batch of 1000 problem instances has been simulated in one run.}
    \label{transitionadapt}
\end{figure}

The results are shown in Fig.~\ref{transitionadapt}. As expected, for low noise, all instances are solved. By increasing the noise strength, the system goes through a sort of ``phase transition'' where the number of successfully solved instances decreases dramatically. As in the case of numerical noise (cf. Fig.~\ref{fig:1}) this solvable-unsolvable transition becomes sharper as the number of variables, $N$, increases. \cite{YuanhangDP} However, also in this case, it is not obvious if such a transition will remain continuous in the
``thermodynamic limit'' ($N\rightarrow\infty$) or will become discontinuous (first order). Such a theoretical analysis is beyond the scope of this paper.

 \subsection{Explanation of the peaks in the spectra} \label{peaks}

 \begin{figure}[t]
\centering
    (a) \includegraphics[width=0.85\columnwidth]{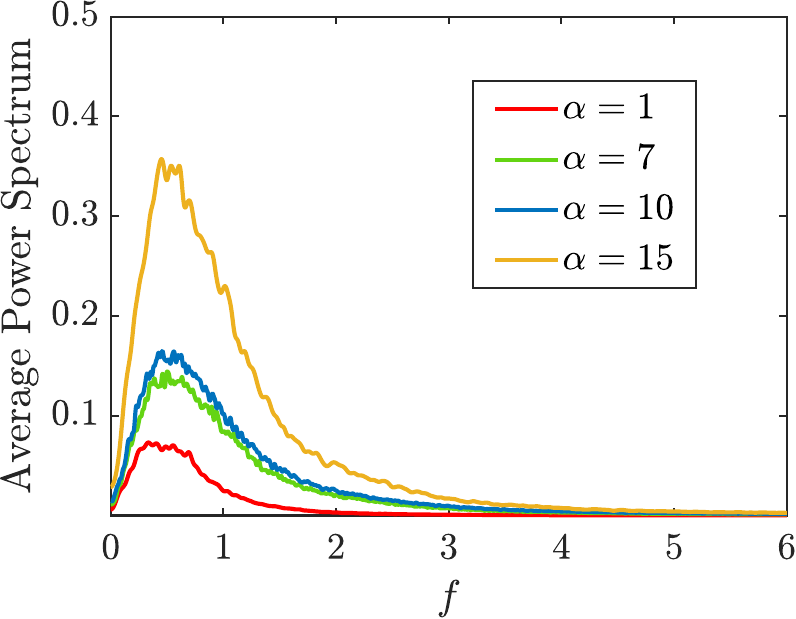}\\
    (b) \includegraphics[width=0.85\columnwidth]{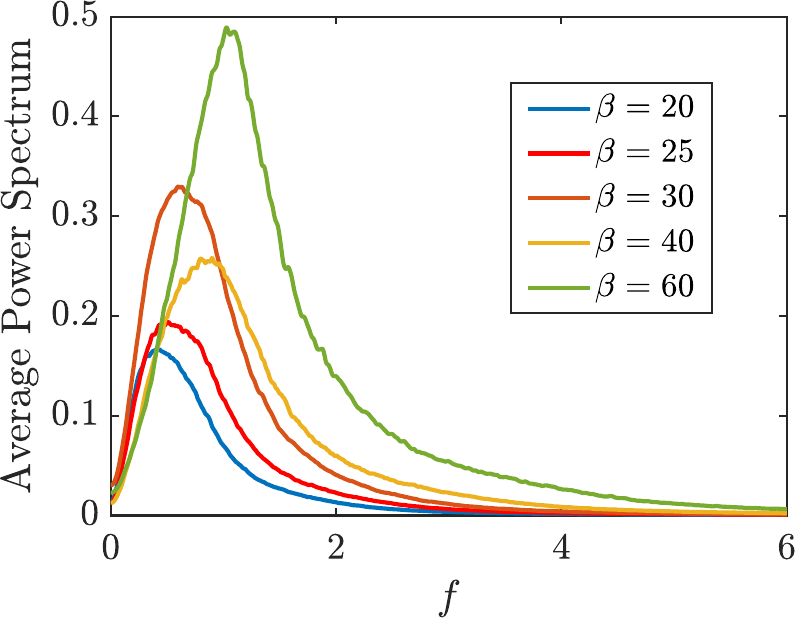}
    \caption{Evolution of the averaged power spectra for different parameter settings in simulations with no noise. The spectra are averaged over 10 different problem instances of $N=1000$, clause-to-variables
    ratio of $M/N=7$, and time step $\Delta t$ of 0.05. (a): Variation of $\alpha$ with fixed $\beta=25$. As $\alpha$ increases, the spectrum peak amplitude grows indicating stronger oscillatory behavior, while the peak frequency remains approximately constant (b): Variation of $\beta$ with fixed $\alpha=5$. The spectrum peak shifts to higher frequencies and its amplitude grows with increasing $\beta$, revealing the system's strong sensitivity to this parameter in controlling oscillation frequency.}
    \label{peaksshift}
\end{figure}

Finally, to understand the origin of the peaks in the power spectra, we refer to Eqs. \eqref{eq:2} and \eqref{eq:3}, in the absence of both physical and numerical noise. These equations show that $\beta$ and $\alpha$ control the rates of change of the memory variables $x_{m}$ and $y_{m}$, respectively. We report in Fig. \ref{peaksshift} simulations performed without physical noise and with all other parameters held constant. The power spectra are averaged over 10 different problem instances. As shown in Fig. \ref{peaksshift}, increasing $\beta$ causes the peak of the power spectrum to shift noticeably toward higher frequencies, whereas increasing $\alpha$ appears to have little to no effect on its position. The difference is that $x_{m}$ appears in its own rate equation, creating a self-reinforcing feedback loop, while $y_{m}$ does not appear in its own rate equation. Thus, $\beta$ controls a state-dependent feedback mechanism in Eq. \eqref{eq:2} inducing faster fluctuations in the variables of Eq. \eqref{eq:1}, where $\beta$ acts indirectly through multiplicative influence on $x_{m}$. In contrast, increasing $\alpha$ has little effect on the spectral peak: while it accelerates the dynamics of $y_{m}$ in Eq. \eqref{eq:3}, this term is not multiplied by any internal state variable and contributes more smoothly to the varaiables $\textnormal{v}_n$'s dynamics. Consequently, its influence does not substantially shift the dominant frequencies in $\dot{\textnormal{v}}_n$.

\section{Discussion} \label{sec:4}

In this study, we employed various indicators to examine how numerical and physical noise impact the behavior of digital memcomputing machines. From a practical perspective, the influence of both types of noise appears similar: with increasing noise strength, the machines' problem-solving capability diminishes from 100\% to a lower percentage, eventually to 0\%. This is evident in Fig.~\ref{fig:1} for numerical noise and Fig.~\ref{transition_n=1000} for physical noise.

To analyze the loss of solvability, we employed the Lyapunov exponent analysis. We utilized the algorithm by Benettin {\it et al.}~\cite{benettin1980lyapunov} in the study of numerical noise, and a conventional algorithm to handle physical noise, leading to the determination of the largest Lyapunov exponents\footnote{We note that the Lyapunov exponent analysis was focused exclusively on the fast degrees of freedom (the logical variables). We believe that these are adequate for assessing whether the overall system dynamics (logical variables combined with memory variables) exhibit chaotic behavior.}. The EAMLLEs are presented in Fig.~\ref{fig:1} for numerical noise and in Fig.~\ref{transition_n=1000} for physical noise.
\textcolor{black}{ With numerical noise, a strong increase in the EAMLLE precedes the decline in the solved-instance percentage, whereas with physical noise a slight delay is observed. }
The reason for this observation is a possible underestimation of LLE by the standard algorithm that evaluates a single exponent versus the spectra in the algorithm by Benettin {\it et al.}~\cite{benettin1980lyapunov}.  Additionally, the standard algorithm was applied to a modified system (without constraints on the $\textnormal{v}_n$'s) that may have a slightly different behavior. Based on the LE analysis, we conclude that the solvable-unsolvable transition in digital memcomputing machines can be associated with a noise-induced chaos.


The power spectra found in this work (see, e.g., Figs.~{\ref{fig:2}}-\ref{transition_n=1000}) provide additional insight into the dynamics of digital memcomputing machines.
In the case of numerical noise, we have noticed that all instances go through four stages when increasing the noise level: at small time steps, they exhibit either no chaos or weak ``transient chaos'' by showing very little or no chaotic behavior; as the time step increases, they enter the stage of ``transient chaos'' while still being able to reach the solution; and by increasing the time step further the system becomes fully chaotic and does not solve anymore. In the final stage, when $\Delta t$ is very large, the power spectra include spikes indicating a strong periodic component. It was observed that at very large time steps, the logical variables oscillate between $\pm 1$. In the case of physical noise, all stages but the last one are also present. This difference may be explained by the fact that unlike physical noise, numerical noise is ``non-local'' in time.

\section{Conclusions} \label{sec:5}

In conclusion, our investigation into the noisy dynamics of digital memcomputing machines (DMMs) demonstrates a fairly high robustness when solving \textcolor{black}{3-SAT } problems in the presence of numerical or physical noise. The effects of noise have been observed through several indicators, such as the percentage of solved instances, power spectra, and Lyapunov exponents. We have observed that the effects of numerical and physical noise are overall qualitatively similar. The largest difference has been found in the case of very strong noise, which, in the numerical case, leads to quasi-periodic-like behavior (in the form of well distinguishable spikes in the power spectra), while strong physical noise results in broadband spectra. Additionally, we emphasize that there exists a ``transiently chaotic'' regime \textcolor{black}{for both types of noise}, characterized by positive mean largest Lyapunov exponents, where DMMs still demonstrate solvability. These results shed further
light on this new class of computing machines and could be used to control whether they operate in the optimal dynamical regime.

\textcolor{black}{
\section*{Supplementary Material}
The supplementary material includes additional power spectra that further illustrate the dynamics of memcomputing machines in the presence of numerical and physical noise.
}

\begin{acknowledgments}
This work was supported by NSF grant No. ECCS-2229880.
\end{acknowledgments}

\renewcommand{\thefigure}{A\arabic{figure}}
\setcounter{figure}{0}

\section*{Data Availability Statement}

The data that support the findings of this study are available from the corresponding author upon reasonable request.

\appendix

\section{Visualization of transient chaos in DMM}\label{Appendix_tran}
\begin{figure}[h]
\centering
    \includegraphics[width=0.85\columnwidth]{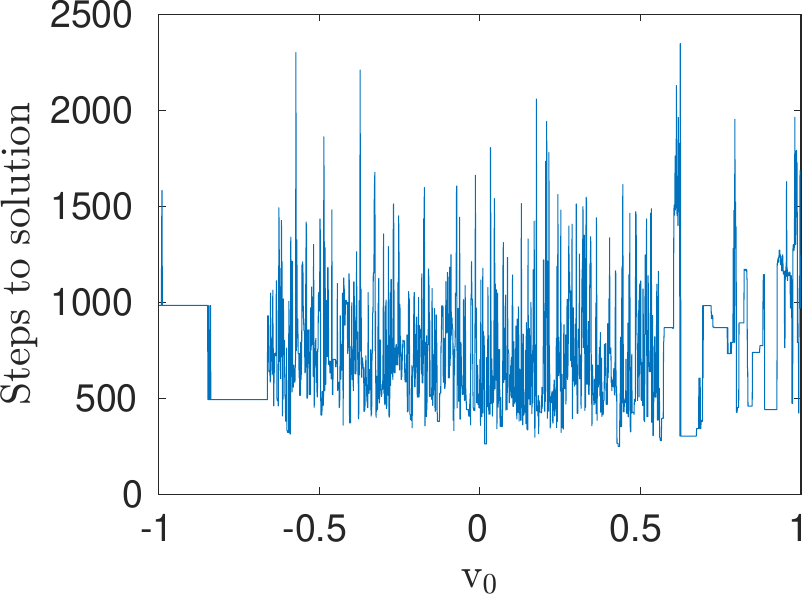}\\
    \caption{\textcolor{black}{Number of integration steps to solution as a function of initial value of $\textnormal{v}_0$ for one instance with $N=1000$ and $M/N=7$ using time step $\Delta t = 0.1$. $\textnormal{v}_0$ is varied monotonically from $-1$ to 1 with an increment of $10^{-3}$.}}
    \label{fig:transient}
\end{figure}

To illustrate the presence of transient chaos in the dynamics of the DMM, we show the number of integration steps required to reach the solution as a function of the initial value $\textnormal{v}_0$ for a single 3-SAT instance in Fig.~\ref{fig:transient}. All other variables were initialized identically and $\textnormal{v}_0$ was continuously varied from -1 to 1.

\section{Time evolution of largest Lyapunov exponents}\label{AppendixA}

\begin{figure}[h]
\centering
    (a) \includegraphics[width=0.85\columnwidth]{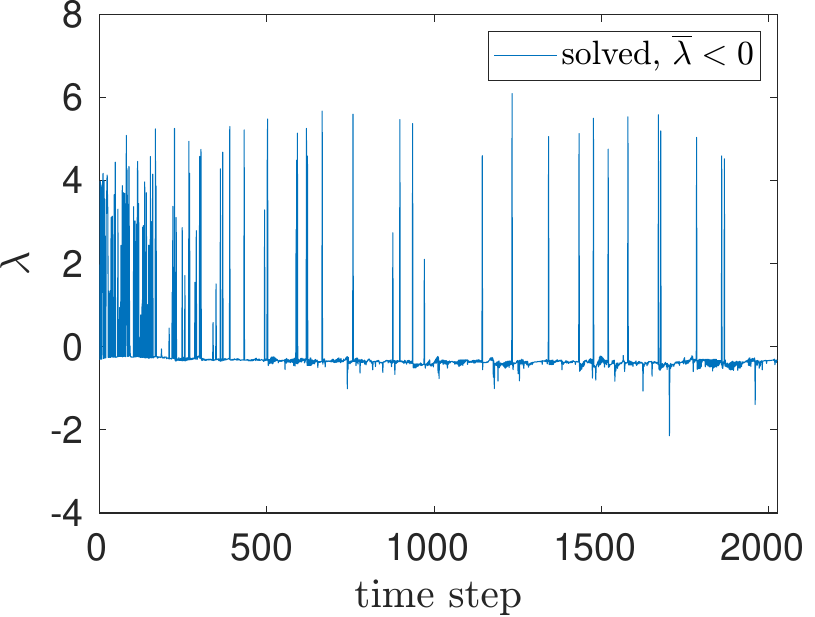}\\
    (b) \includegraphics[width=0.85\columnwidth]{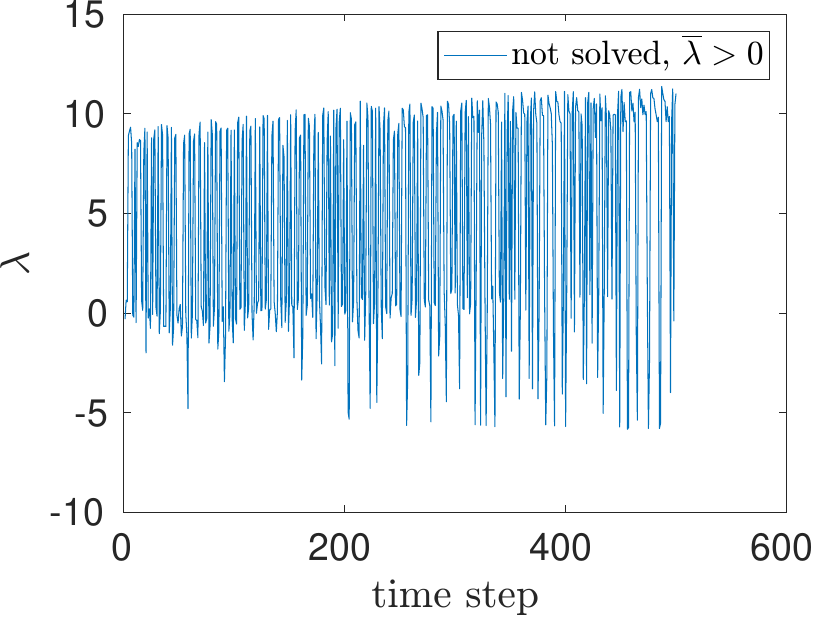}
    \caption{Largest Lyapunov exponent, LLE, as a function of time for the same instance as in  Fig.~S1 calculated using the time step of (a) $0.02$ and (b) $0.36$.}
    \label{fig:lyap_example}
\end{figure}

In Fig. ~\ref{fig:lyap_example}, we plot the LLE as a function of time throughout the dynamics for cases with both small and large numerical noise. The trajectories exhibit large spikes in LLE, which we attribute to the existence of saddle \textcolor{black}{ orbits } in the phase space. The unstable directions of the saddle \textcolor{black}{ orbits } would naturally produce positive LLEs.

\section{$1/f$ noise}\label{1/f_noise}
As mentioned in the main text, in the case of physical noise there is an additional feature present in the power spectra at very low frequencies (Inset (2) in Fig. \ref{transition_n=1000}). We attribute this feature to flicker noise~\cite{DuttaHorn}. This phenomenon is characterized by a power spectral density that scales inversely with frequency, following the form $1/f^{\alpha}$, with typically $1\leq\alpha\leq2$.

\begin{figure}[tb]
\centering
    (a) \includegraphics[width=0.85\columnwidth]{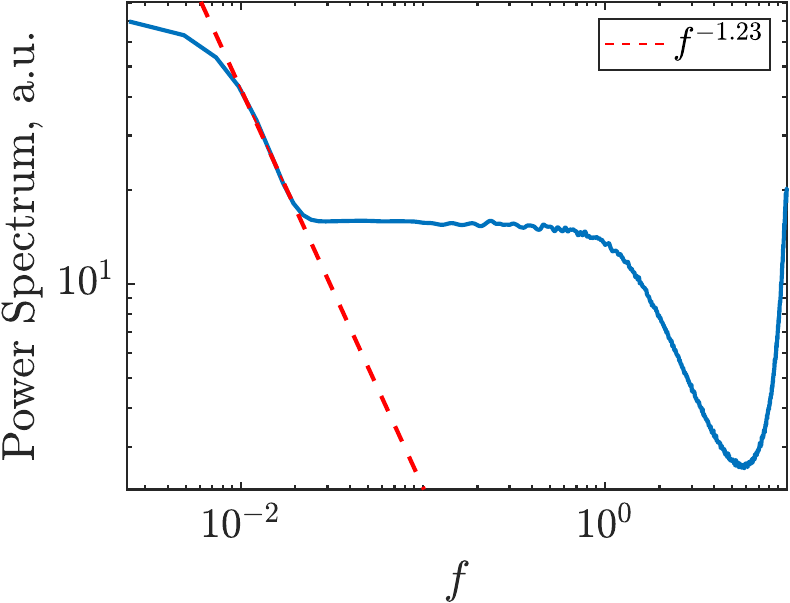} \\
    (b) \includegraphics[width=0.85\columnwidth]{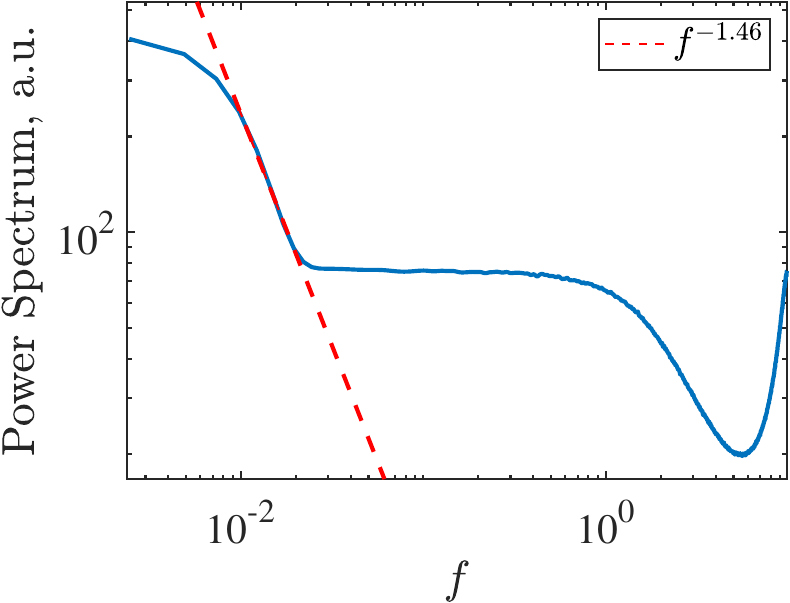}
    \caption{Log-log plot of the power spectrum for the time derivative of the $\textnormal{v}_n$'s. We simulated a problem instance with $N=1000$, with ratio $M/N=7$. In any stochastic process, there is a $1/f$ power spectral density down to the lowest frequency $f$ whenever $f>\Omega_{min}$, with $\Omega_{min}$ being the minimum relaxation frequency~\cite{Diaz_2014}.  Since the minimum relaxation frequency of a system is generally fixed, the power spectrum deviates from the $1/f$ trend for frequencies $f$ smaller than $\Omega_{min}$, marking a crossover to a different spectral regime. On (a), the power spectrum is averaged over 100 realizations with noise strength $\Gamma=1$, while for (b) the noise strength is $\Gamma=2$ and the same average is performed.}
    \label{linearfit}
\end{figure}

This is illustrated in the power spectra of Fig. \ref{transition_n=1000}. The characteristic
$1/f$ behavior is confirmed by the spectral slopes shown in Fig. \ref{linearfit}, observed for both intermediate and high noise strength. Indeed, at low frequencies, the spectrum scales as $f^{-1.23}$ for $\Gamma=1$ and as $f^{-1.46}$ for $\Gamma=2$. The noise
saturates at frequencies comparable to the lowest relaxation frequency of the process~\cite{Diaz_2014}.

Apparently, numerical noise also contributes to the appearance of $1/f$ noise observed in certain power spectra. An illustration can be found in Fig.~S3 of the Supplementary Information (SI). However, this observation is instance/realization dependent. In particular, there is no visible $1/f$ noise in the power spectra in Fig.~S2(a)-(c) in the SI.

%

\clearpage
\newpage

\begin{widetext}

\centering \Large{Supplementary Information for the manuscript:\\ On the solvable–unsolvable transition
due to noise-induced chaos in digital memcomputing}

\renewcommand\thefigure{S\arabic{figure}}
\renewcommand{\thepage}{S\arabic{page}}

\setcounter{figure}{0}
\setcounter{page}{1}

\begin{figure}[h]
\centering
    (a) \includegraphics[width=0.4\textwidth]{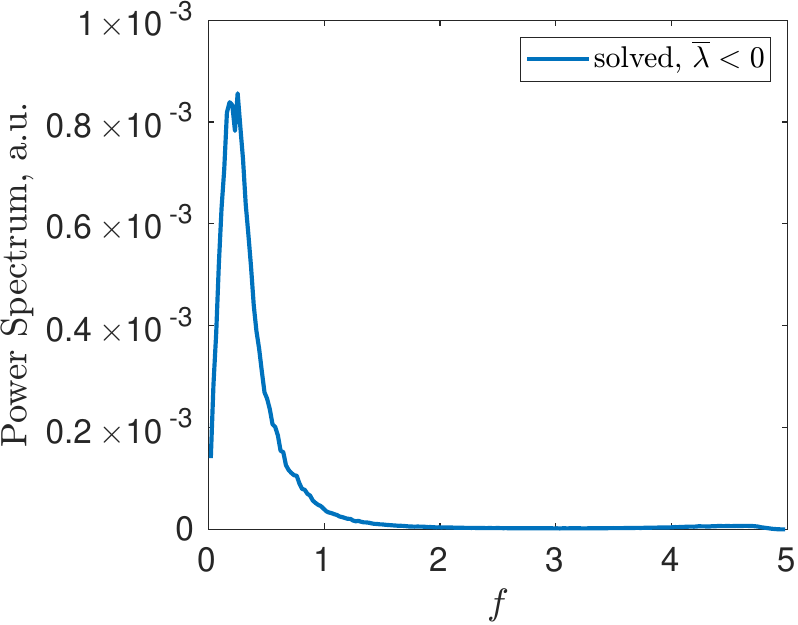}
    (b) \includegraphics[width=0.4\textwidth] {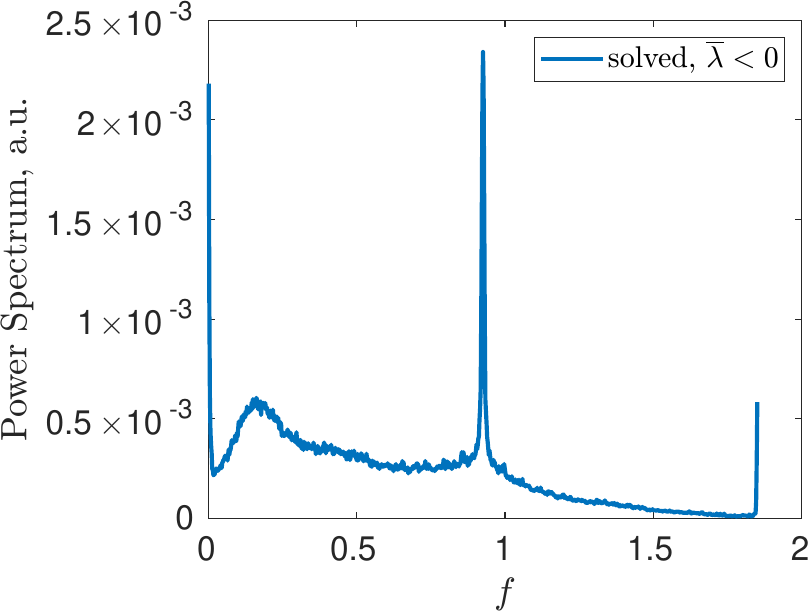} \\
    (c) \includegraphics[width=0.4\textwidth]{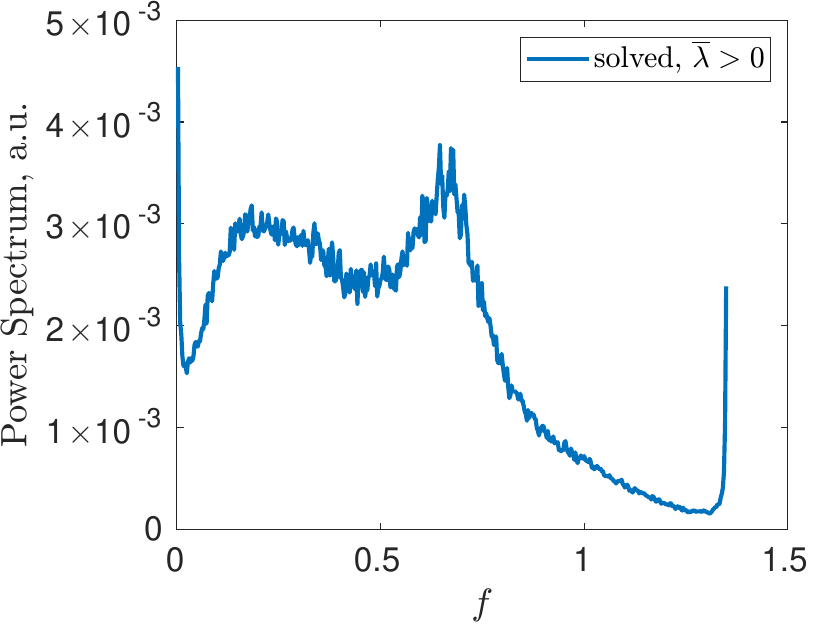}
    (d) \includegraphics[width=0.4\textwidth] {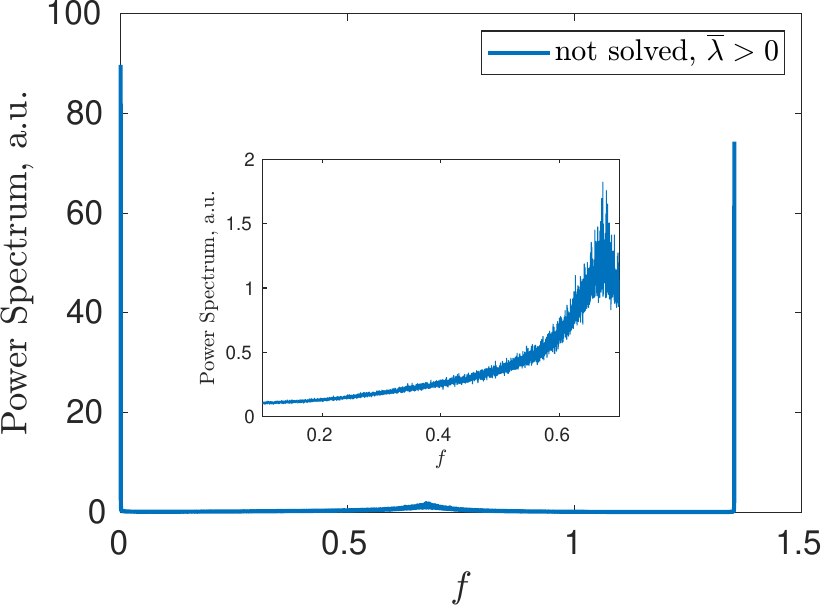}
    \caption{Power spectra for a satisfiable instance with 1000 variables and  with clause-to-variable ratio $M/N=7$ generated using the integration time step of (a) 0.1, (b) 0.27, (c) 0.35, and (d) 0.37. We note that these curves are typical for the instances with $M/N=7$ except for the sharp peak in (b) at $f\approx 1$ and the broad peak at $f\approx 0.7$ in (c). Additionally, the curves in (b) and (c) exhibit a characteristic $1/f$ noise behavior at very low frequencies, which we discuss in Appendix B.}
    \label{fig:PS_example_1}
\end{figure}

\begin{figure}[h]
\centering
    (a) \includegraphics[width=0.4\textwidth]{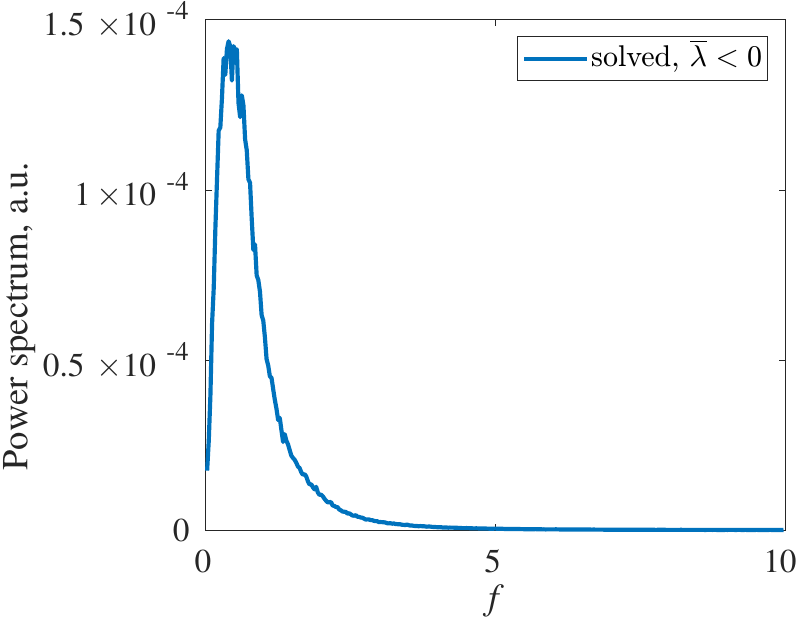}     (b) \includegraphics[width=0.4\textwidth]{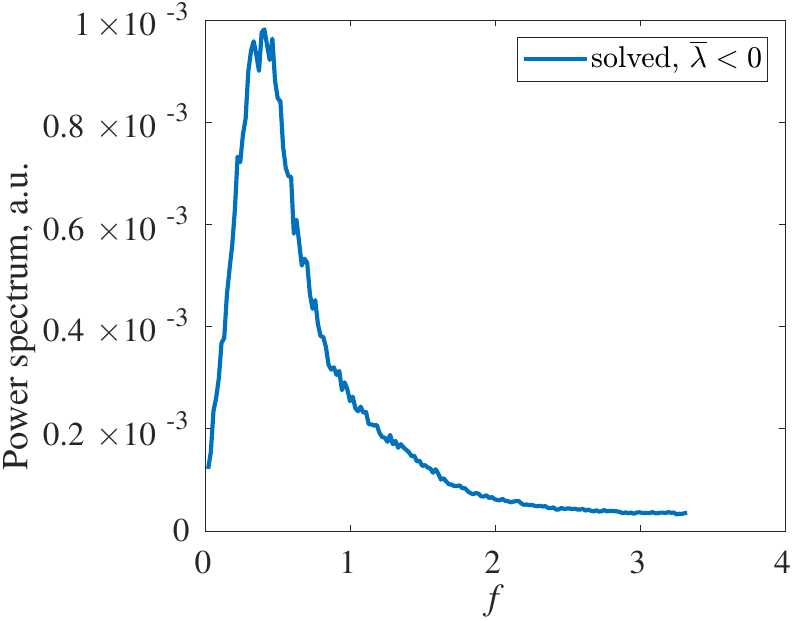} \\
    (c) \includegraphics[width=0.4\textwidth]{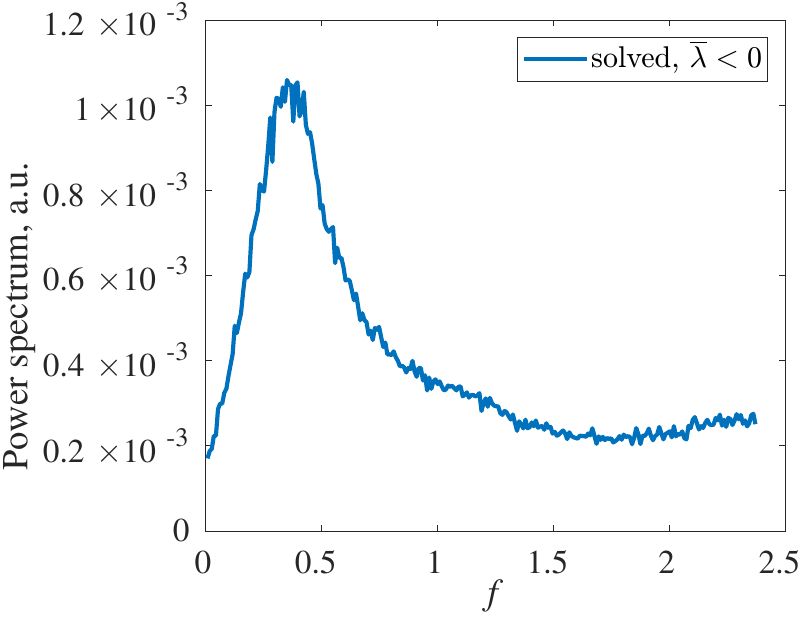}(d) \includegraphics[width=0.4\textwidth]{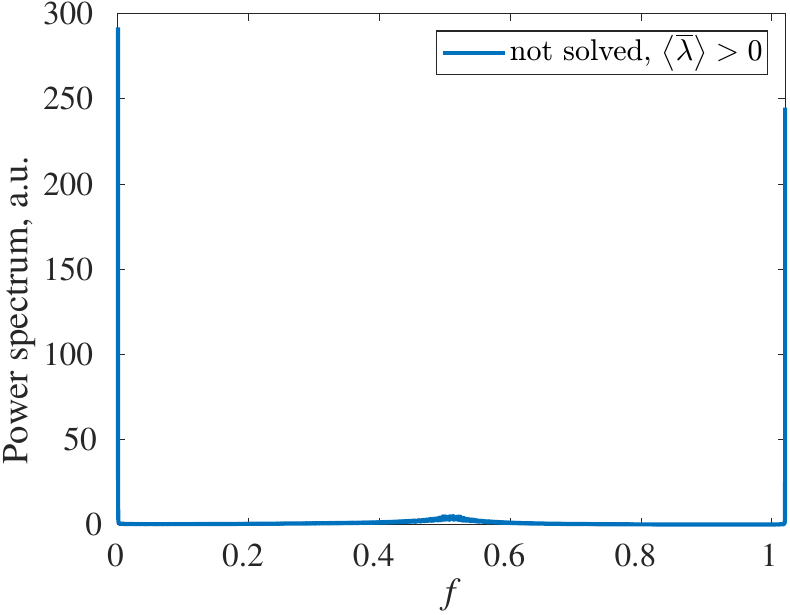} \\
    \caption{Power spectra for a satisfiable instance with 1000 variables and  with clause-to-variable ratio $M/N=7$ generated using the integration time step of (a) 0.05, (b) 0.15, (c) 0.21, and (d) 0.49.}
    \label{fig:PS_example_2}
\end{figure}

\begin{figure}[h]
\centering
\includegraphics[width=0.4\textwidth]{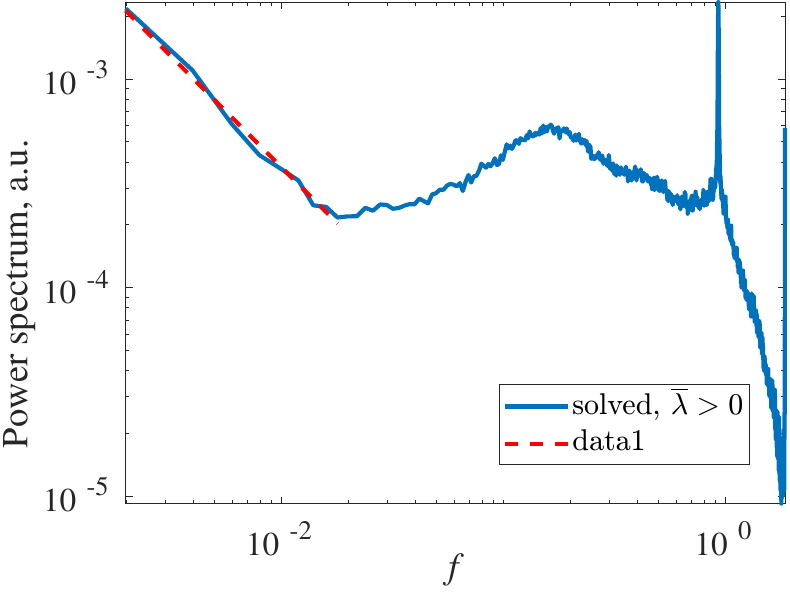}
\caption{Log-log plot of Fig.~\ref{fig:PS_example_1}(b). The fitting curve (``data1'') is $\propto 1/f^{1.07}$. The curve in Fig.~\ref{fig:PS_example_1}(c) follows a similar $1/f$ behavior with a slightly different exponent. In Fig.~\ref{fig:PS_example_1}(d), however, the  spike at $f=0$ consists of a single point.}
    \label{fig:PS_example_11}
\end{figure}

\begin{figure}[h]
\centering
    (a) \includegraphics[width=0.4\textwidth]{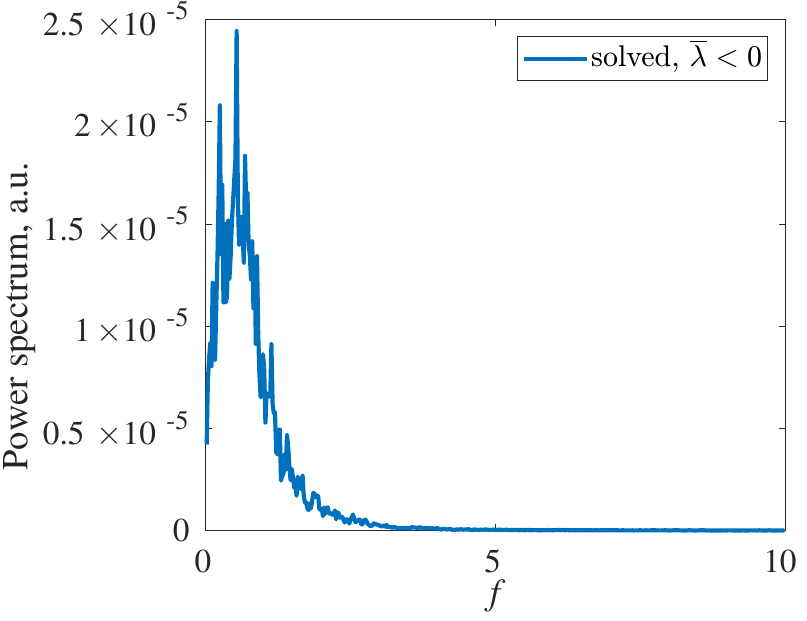}
    (b) \includegraphics[width=0.4\textwidth]{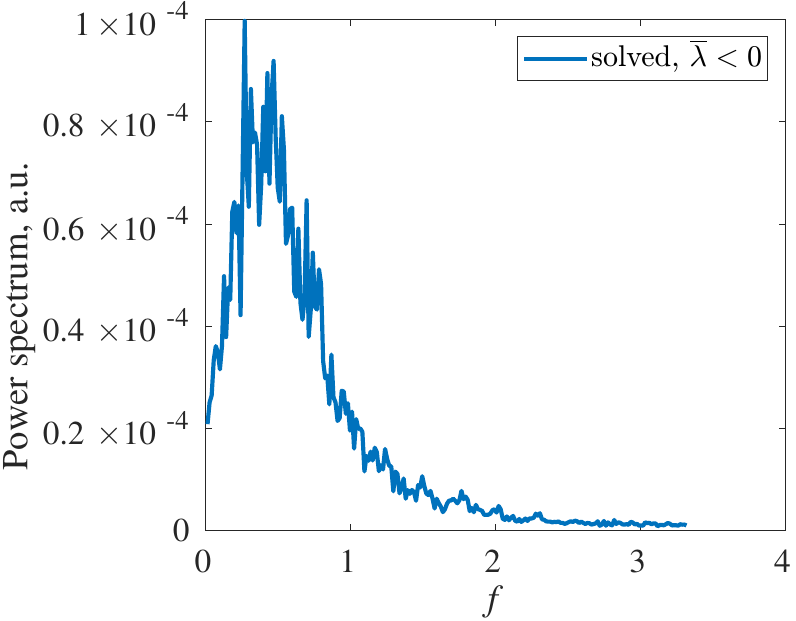} \\
    (c) \includegraphics[width=0.4\textwidth]{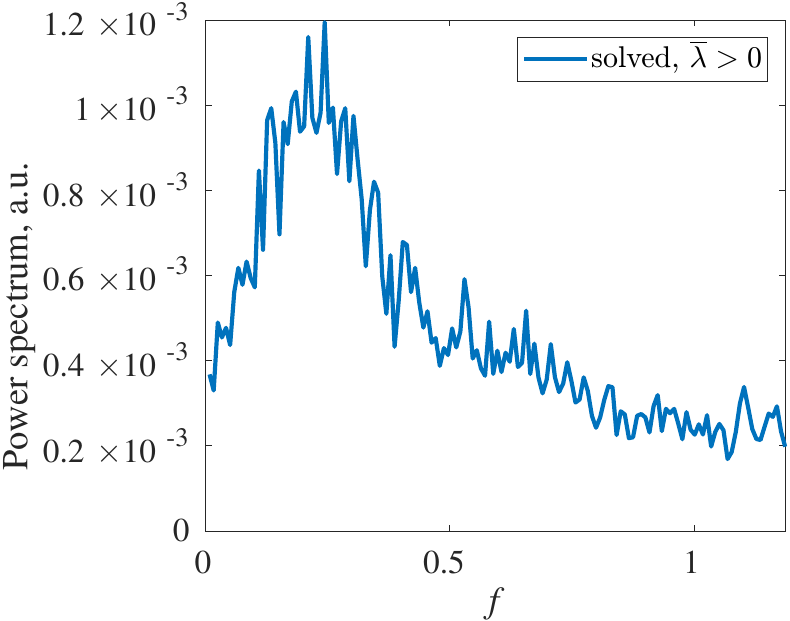}
    (d) \includegraphics[width=0.4\textwidth]{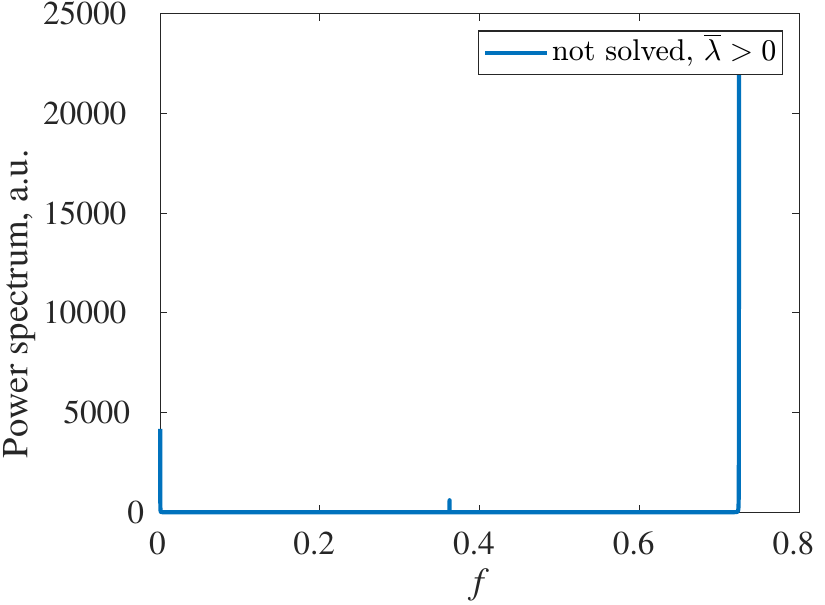}
    \caption{Power spectra for a satisfiable instance with 90 variables and  with clause-to-variable ratio $M/N=4.3$ generated using the integration time step of (a) 0.05, (b) 0.15, (c) 0.42, and (d) 0.69.}
    \label{fig:PS_example_3}
\end{figure}

\begin{figure}[h]
    (a) \includegraphics[width=0.4\textwidth]{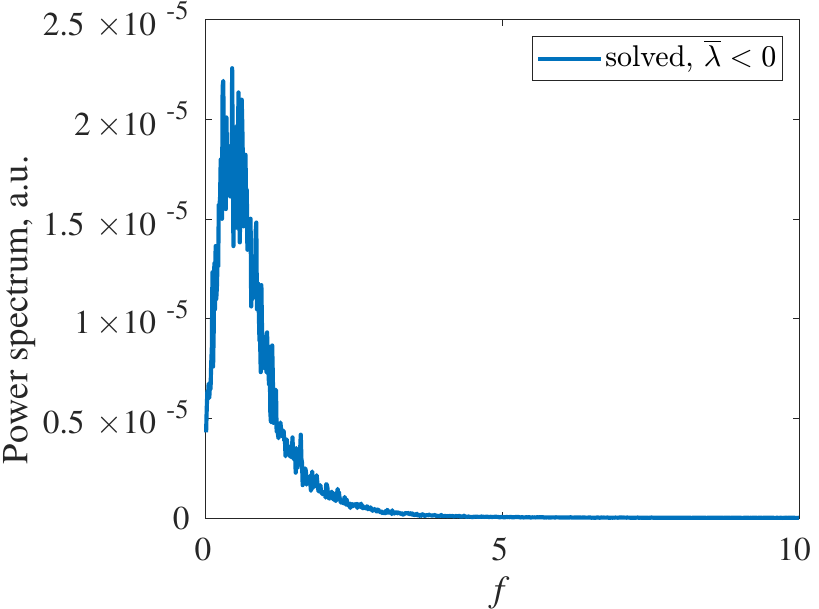}
    (b) \includegraphics[width=0.4\textwidth]{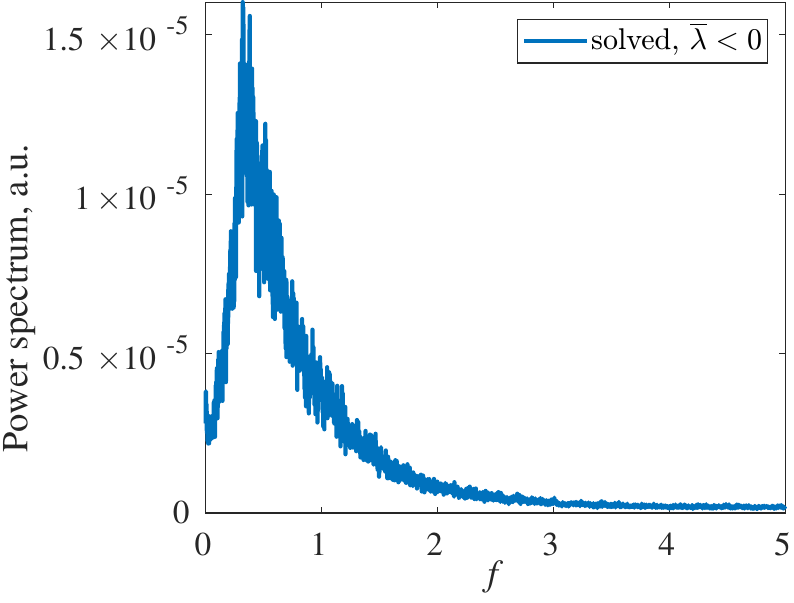}\\
    (c) \includegraphics[width=0.4\textwidth]{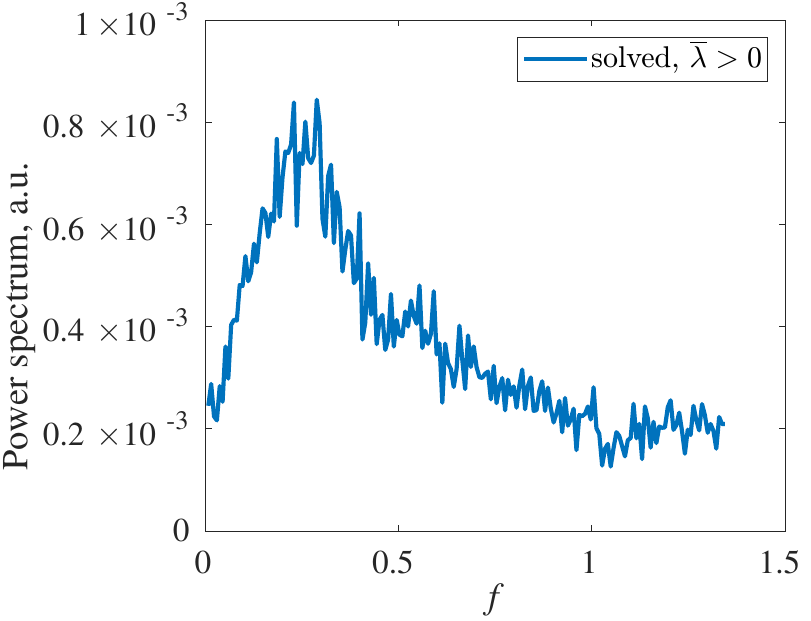}
    (d) \includegraphics[width=0.4\textwidth]{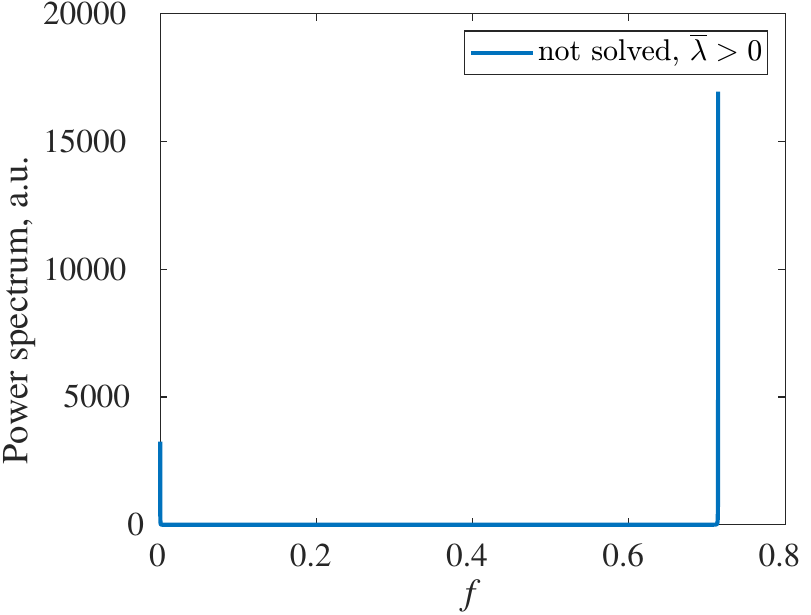}
    \caption{Power spectra for a satisfiable instance with 90 variables and  with clause-to-variable ratio $M/N=4.3$ generated using the integration time step of (a) 0.05, (b) 0.1, (c) 0.37, and (d) 0.7.}
    \label{fig:PS_example_4}
\end{figure}

\begin{figure}
\centering
    (a) \includegraphics[width=0.4\textwidth]{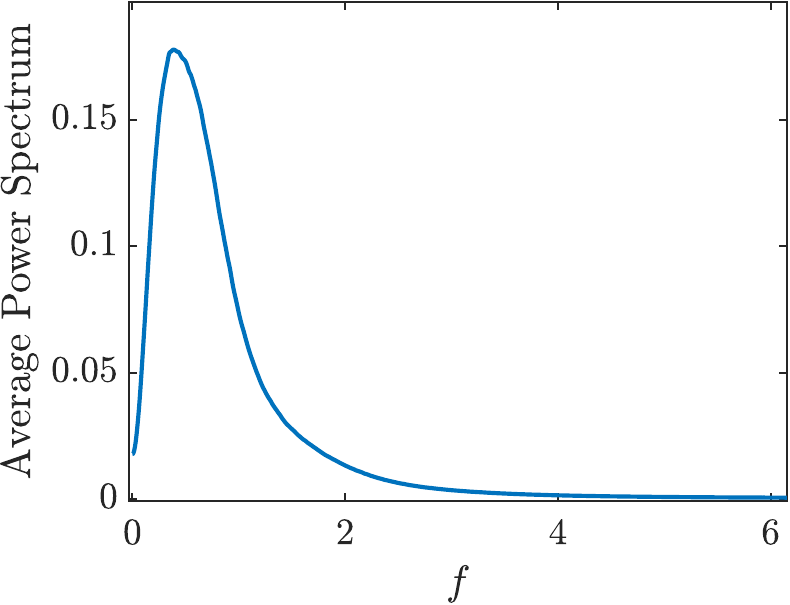}
    (b) \includegraphics[width=0.4\textwidth]{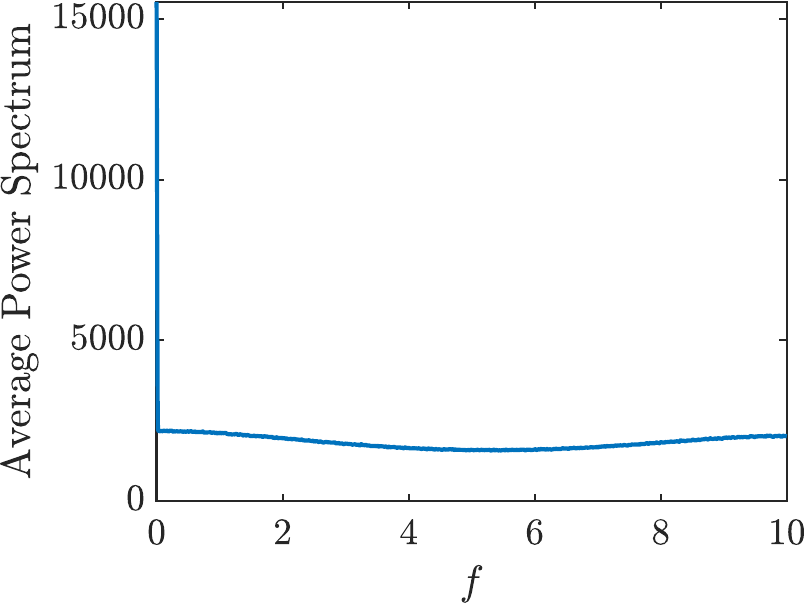}
    \caption{Panel (a) shows the average power spectrum of 100 randomly chosen problem instance with $N=1000$, $M/N=7$, and random initial $v_n$'s in the absence of noise. Panel (b) shows the averaged power spectrum of 100 problem instances with $N=1000$, $M/N=7$, and with noise strength of $\Gamma = 2$. None of these instances were solved. In both panels, the power spectra exhibit similar characteristics of those observed in Fig. 4: the peak in Panel (a) appears at the same frequency, and in Panel (b) the spectrum (with noise), shows an increase in power at high frequencies, attributed to long-term memory effects.}
    \label{randominstances}
\end{figure}

\end{widetext}

\end{document}